%% file: main.tex
\pdfoutput=1
\documentclass[12pt,a4paper]{article}

\usepackage{ifthen}

\usepackage{slashbox}
\usepackage{rotating}
\usepackage{multirow}
\usepackage{graphpap}

\newboolean{pdflatex}
\setboolean{pdflatex}{true}

\newboolean{articletitles}
\setboolean{articletitles}{true}

\newboolean{uprightparticles}
\setboolean{uprightparticles}{true}

\input{local-symbols}

\input{preamble}
\begin{document}

\input{title-LHCb-PAPER}


\pagestyle{plain}
\setcounter{page}{1}
\pagenumbering{arabic}

\cleardoublepage
\input{intro_combined}

\input{detector_combined}

\input{evsel_combined}

\input{ana2_combined}

\input{ana_combined}

\input{ana3_combined}

\input{ratios_combined}

\input{summary_combined}

\input{acknowledgements}

\addcontentsline{toc}{section}{References}
\bibliographystyle{LHCb}
\bibliography{main,local}

\end{document}

%% file: local-symbols.tex
\def\Psigma  {\ensuremath{\upsigma}\xspace}                 

\def\pipipi  {\ensuremath{\pipi{}\piz}\xspace}
\def\gamgam  {\ensuremath{\g{}\g}\xspace}

%% file: preamble.tex
\usepackage{microtype}
\usepackage{lineno}  
\usepackage{xspace} 

\usepackage{graphicx}  
\usepackage{color}
\usepackage{colortbl}
\graphicspath{{./figs/}} 

\usepackage{amsmath} 
\usepackage{amssymb}
\usepackage{amsfonts}
\usepackage{upgreek} 

\newcommand*\patchAmsMathEnvironmentForLineno[1]{%
\expandafter\let\csname old#1\expandafter\endcsname\csname #1\endcsname
\expandafter\let\csname oldend#1\expandafter\endcsname\csname
end#1\endcsname
 \renewenvironment{#1}%
   {\linenomath\csname old#1\endcsname}%
   {\csname oldend#1\endcsname\endlinenomath}%
}
\newcommand*\patchBothAmsMathEnvironmentsForLineno[1]{%
  \patchAmsMathEnvironmentForLineno{#1}%
  \patchAmsMathEnvironmentForLineno{#1*}%
}
\AtBeginDocument{%
\patchBothAmsMathEnvironmentsForLineno{equation}%
\patchBothAmsMathEnvironmentsForLineno{align}%
\patchBothAmsMathEnvironmentsForLineno{flalign}%
\patchBothAmsMathEnvironmentsForLineno{alignat}%
\patchBothAmsMathEnvironmentsForLineno{gather}%
\patchBothAmsMathEnvironmentsForLineno{multline}%
}

\usepackage{hyperref}    
\usepackage[all]{hypcap} 

\input{lhcb-symbols-def} 

\usepackage{cite} 
\usepackage{mciteplus}

%% file: lhcb-symbols-def.tex
\def\lhcb {\mbox{LHCb}\xspace}
\def\ux85 {\mbox{UX85}\xspace}

\ifthenelse{\boolean{uprightparticles}}%
{
 
 \def\Pgamma      {\ensuremath{\upgamma}\xspace}

 \def\Peta        {\ensuremath{\upeta}\xspace}

 \def\Pmu         {\ensuremath{\upmu}\xspace}

 \def\Ppi         {\ensuremath{\uppi}\xspace}                 
                  
 \def\Prho        {\ensuremath{\uprho}\xspace}                 
                  
 \def\Ptau        {\ensuremath{\uptau}\xspace}                 
                  
 \def\Pphi        {\ensuremath{\upphi}\xspace}                 
                  
 \def\Pchi        {\ensuremath{\upchi}\xspace}                 
 \def\Ppsi        {\ensuremath{\uppsi}\xspace}                 
 \def\Pomega      {\ensuremath{\upomega}\xspace}                 

 \def\PDelta      {\ensuremath{\Delta}\xspace}                 
 \def\PXi      {\ensuremath{\Xi}\xspace}                 
 \def\PLambda      {\ensuremath{\Lambda}\xspace}                 
 \def\PSigma      {\ensuremath{\Sigma}\xspace}                 
 \def\POmega      {\ensuremath{\Omega}\xspace}                 
 \def\PUpsilon      {\ensuremath{\Upsilon}\xspace}                 
 
 \def\PB      {\ensuremath{\mathrm{B}}\xspace}                 
                  
 \def\PD      {\ensuremath{\mathrm{D}}\xspace}

 \def\PJ      {\ensuremath{\mathrm{J}}\xspace}                 
 \def\PK      {\ensuremath{\mathrm{K}}\xspace}

 \def\Pb      {\ensuremath{\mathrm{b}}\xspace}                 
 \def\Pc      {\ensuremath{\mathrm{c}}\xspace}

 \def\Pi      {\ensuremath{\mathrm{i}}\xspace}

 \def\Ps      {\ensuremath{\mathrm{s}}\xspace}

}
{
 
 \def\Pgamma      {\ensuremath{\gamma}\xspace}

 \def\Peta        {\ensuremath{\eta}\xspace}

 \def\Pmu         {\ensuremath{\mu}\xspace}

 \def\Ppi         {\ensuremath{\pi}\xspace}                 
                  
 \def\Prho        {\ensuremath{\rho}\xspace}                 
                  
 \def\Ptau        {\ensuremath{\tau}\xspace}                 
                  
 \def\Pphi        {\ensuremath{\phi}\xspace}                 
                  
 \def\Pchi        {\ensuremath{\chi}\xspace}                 
 \def\Ppsi        {\ensuremath{\psi}\xspace}                 
 \def\Pomega      {\ensuremath{\omega}\xspace}                 
 \mathchardef\PDelta="7101
 \mathchardef\PXi="7104
 \mathchardef\PLambda="7103
 \mathchardef\PSigma="7106
 \mathchardef\POmega="710A
 \mathchardef\PUpsilon="7107
                  
 \def\PB      {\ensuremath{B}\xspace}                 
                  
 \def\PD      {\ensuremath{D}\xspace}

 \def\PJ      {\ensuremath{J}\xspace}                 
 \def\PK      {\ensuremath{K}\xspace}

 \def\Pb      {\ensuremath{b}\xspace}                 
 \def\Pc      {\ensuremath{c}\xspace}

 \def\Pi      {\ensuremath{i}\xspace}

 \def\Ps      {\ensuremath{s}\xspace}

}

\def\mumu       {\ensuremath{\Pmu^+\Pmu^-}\xspace}

\def\g      {\ensuremath{\Pgamma}\xspace}

\def\squark    {\ensuremath{\Ps}\xspace}

\def\cquark    {\ensuremath{\Pc}\xspace}

\def\bquark    {\ensuremath{\Pb}\xspace}

\def\pion  {\ensuremath{\Ppi}\xspace}
\def\piz   {\ensuremath{\pion^0}\xspace}

\def\pipi  {\ensuremath{\pion^+\pion^-}\xspace}

\def\kaon  {\ensuremath{\PK}\xspace}
  \def\Kbar  {\kern 0.2em\overline{\kern -0.2em \PK}{}\xspace}

\def\Kz    {\ensuremath{\kaon^0}\xspace}
\def\Kzb   {\ensuremath{\Kbar^0}\xspace}
\def\KzKzb {\ensuremath{\Kz \kern -0.16em \Kzb}\xspace}
\def\Kp    {\ensuremath{\kaon^+}\xspace}
\def\Km    {\ensuremath{\kaon^-}\xspace}

\def\KpKm  {\ensuremath{\Kp \kern -0.16em \Km}\xspace}

  \def\Dbar    {\kern 0.2em\overline{\kern -0.2em \PD}{}\xspace}
\def\D       {\ensuremath{\PD}\xspace}

\def\Dz      {\ensuremath{\D^0}\xspace}
\def\Dzb     {\ensuremath{\Dbar^0}\xspace}
\def\DzDzb   {\ensuremath{\Dz {\kern -0.16em \Dzb}}\xspace}
\def\Dp      {\ensuremath{\D^+}\xspace}
\def\Dm      {\ensuremath{\D^-}\xspace}

\def\DpDm    {\ensuremath{\Dp {\kern -0.16em \Dm}}\xspace}

\def\B       {\ensuremath{\PB}\xspace}
  \def\Bbar    {\kern 0.18em\overline{\kern -0.18em \PB}{}\xspace}

\def\Bd      {\ensuremath{\B^0}\xspace}
\def\Bs      {\ensuremath{\B^0_\squark}\xspace}

\def\jpsi     {\ensuremath{{\PJ\mskip -3mu/\mskip -2mu\Ppsi\mskip 2mu}}\xspace}

  \def\Y#1S{\ensuremath{\PUpsilon{(#1S)}}\xspace}

\def\Lbar {\ensuremath{\kern 0.1em\overline{\kern -0.1em\PLambda}}\xspace}

\def\BF         {{\ensuremath{\cal B}\xspace}}

\def\BR         {\BF}

\def\to                 {\ensuremath{\rightarrow}\xspace}

\def\CP                {\ensuremath{C\!P}\xspace}

\def\AT#1     {\ensuremath{A_{\mathrm{T}}^{#1}}\xspace}           

\def\C#1      {\ensuremath{\mathcal{C}_{#1}}\xspace}                       
\def\Cp#1     {\ensuremath{\mathcal{C}_{#1}^{'}}\xspace}                    
\def\Ceff#1   {\ensuremath{\mathcal{C}_{#1}^{\mathrm{(eff)}}}\xspace}        
\def\Cpeff#1  {\ensuremath{\mathcal{C}_{#1}^{'\mathrm{(eff)}}}\xspace}       
\def\Ope#1    {\ensuremath{\mathcal{O}_{#1}}\xspace}                       
\def\Opep#1   {\ensuremath{\mathcal{O}_{#1}^{'}}\xspace}                    

\newcommand{\tev}{\ensuremath{\mathrm{\,Te\kern -0.1em V}}\xspace}
\newcommand{\gev}{\ensuremath{\mathrm{\,Ge\kern -0.1em V}}\xspace}
\newcommand{\mev}{\ensuremath{\mathrm{\,Me\kern -0.1em V}}\xspace}
\newcommand{\kev}{\ensuremath{\mathrm{\,ke\kern -0.1em V}}\xspace}
\newcommand{\ev}{\ensuremath{\mathrm{\,e\kern -0.1em V}}\xspace}
\newcommand{\gevc}{\ensuremath{{\mathrm{\,Ge\kern -0.1em V\!/}c}}\xspace}
\newcommand{\mevc}{\ensuremath{{\mathrm{\,Me\kern -0.1em V\!/}c}}\xspace}
\newcommand{\gevcc}{\ensuremath{{\mathrm{\,Ge\kern -0.1em V\!/}c^2}}\xspace}
\newcommand{\gevgevcccc}{\ensuremath{{\mathrm{\,Ge\kern -0.1em V^2\!/}c^4}}\xspace}
\newcommand{\mevcc}{\ensuremath{{\mathrm{\,Me\kern -0.1em V\!/}c^2}}\xspace}

\def\mum  {\ensuremath{\,\upmu\rm m}\xspace}

\def\gsim{{~\raise.15em\hbox{$>$}\kern-.85em
          \lower.35em\hbox{$\sim$}~}\xspace}
\def\lsim{{~\raise.15em\hbox{$<$}\kern-.85em
          \lower.35em\hbox{$\sim$}~}\xspace}

\def\pt         {\mbox{$p_{\rm T}$}\xspace}

\def\evtgen     {\mbox{\textsc{EvtGen}}\xspace}
\def\pythia     {\mbox{\textsc{Pythia}}\xspace}

\def\geant      {\mbox{\textsc{Geant4}}\xspace}

\def\photos     {\mbox{\textsc{Photos}}\xspace}

\def\tell1  {TELL1\xspace}
\def\ukl1   {UKL1\xspace}



%% file: title-LHCb-PAPER.tex

\begin{titlepage}
\pagenumbering{roman}

\vspace*{-1.5cm}
\centerline{\large EUROPEAN ORGANIZATION FOR NUCLEAR RESEARCH (CERN)}
\vspace*{1.5cm}
\hspace*{-0.5cm}
\begin{tabular*}{\linewidth}{lc@{\extracolsep{\fill}}r}
\ifthenelse{\boolean{pdflatex}}
{\vspace*{-2.7cm}\mbox{\!\!\!\includegraphics[width=.14\textwidth]{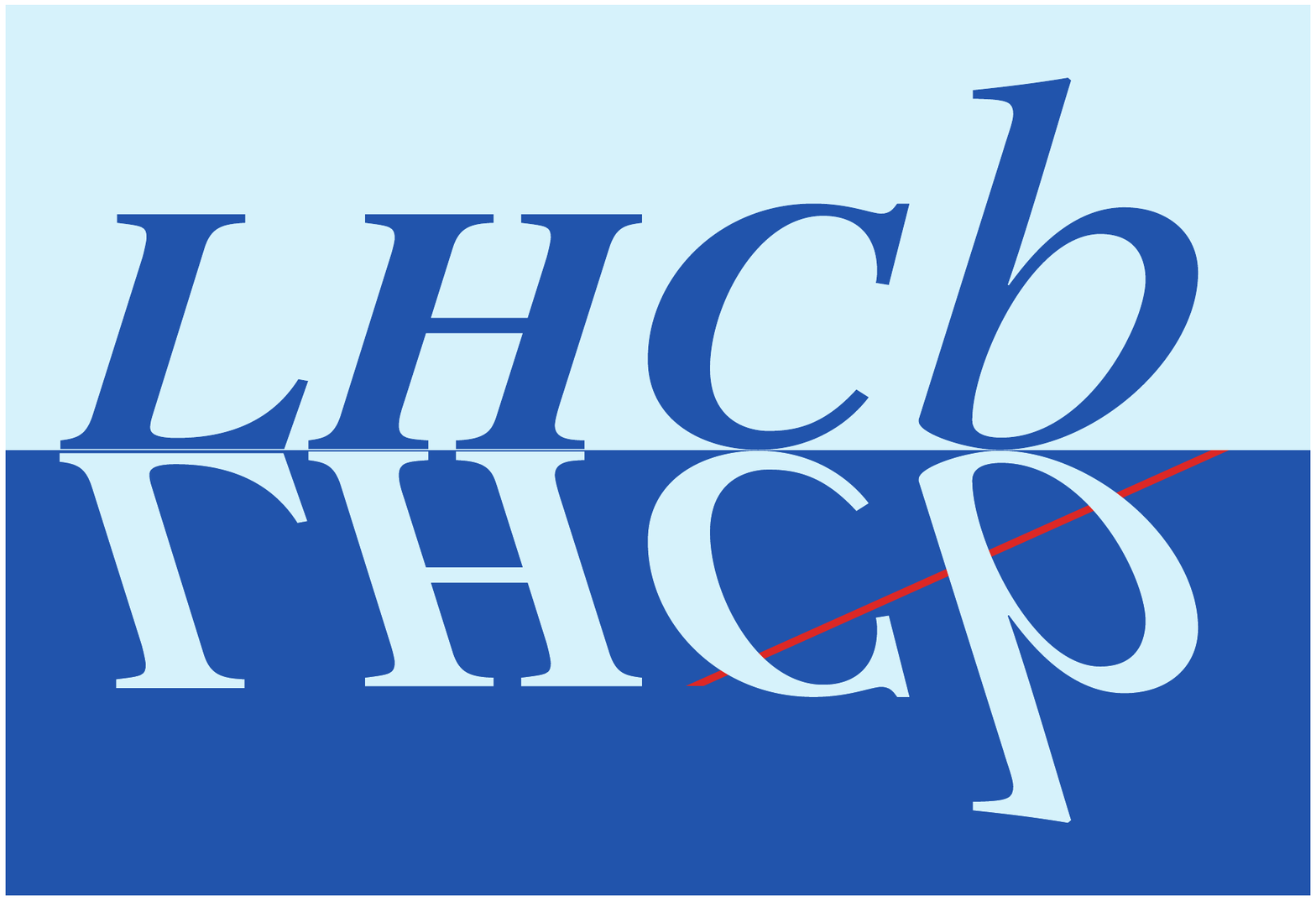}} & &}%
{\vspace*{-1.2cm}\mbox{\!\!\!\includegraphics[width=.12\textwidth]{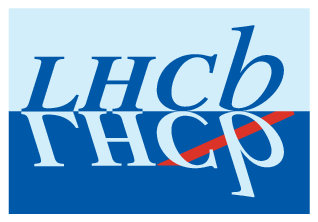}} & &}%
\\
 & & CERN-PH-EP-2012-287 \\  
 & & LHCb-PAPER-2012-022 \\  
 & & \today \\ 
 & & \\
\end{tabular*}

\vspace*{1.2cm}

{\bf\boldmath\huge
\begin{center}
  Evidence for the decay $\mathrm{B}^0\to \jpsi \Pomega$ and measurement 
of the relative branching fractions of $\Bs$ meson decays to $\jpsi\Peta$ and $\jpsi\Peta^{\prime}$
\end{center}
}

\vspace*{0.5cm}

\begin{center}
The LHCb collaboration\footnote{Authors are listed on the following pages.}
\end{center}

\vspace{\fill}

\begin{abstract}
  \noindent
  First evidence of the $\Bd\to\jpsi\Pomega$ decay is found and the $\Bs\to\jpsi\Peta$ and $\Bs\to\jpsi\Peta^{\prime}$ decays
  are studied using a dataset corresponding to an integrated luminosity of 1.0~$\mathrm{fb}^{-1}$ collected by the 
  LHCb experiment in proton-proton 
  collisions at a centre-of-mass energy of $\sqrt{s} = 7$~TeV. The branching fractions of these decays are measured 
  relative to that of the $\Bd\to\jpsi\Prho^0$ decay:
  \begin{equation*}
  \begin{array}{lll}  
  \frac{\BR(\Bd\to\jpsi\Pomega)}{\BR(\Bd\to\jpsi\Prho^0)} $=$ \:0.89 \pm0.19\,(\mathrm{stat})\,^{+0.07}_{-0.13}\,(\mathrm{syst}),
  \\
  \noalign{\vskip 3pt}
  \frac{\BR(\Bs\to\jpsi\Peta)}{\BR(\Bd\to\jpsi\Prho^0)} $=$ \:14.0 \pm 1.2\,(\mathrm{stat})\,^{+1.1}_{-1.5}\,(\mathrm{syst})\,^{+1.1}_{-1.0}\left(\frac{f_\mathrm{d}}{f_\mathrm{s}}\right),
  \\
  \noalign{\vskip 3pt}
  \frac{\BR(\Bs\to\jpsi\Peta^{\prime})}{\BR(\Bd\to\jpsi\Prho^0)} $=$ \:12.7\pm1.1\,(\mathrm{stat})\,^{+0.5}_{-1.3}\,(\mathrm{syst})\,^{+1.0}_{-0.9}\left(\frac{f_\mathrm{d}}{f_\mathrm{s}}\right),
  \end{array}
  \end{equation*} 
  where the last uncertainty is due to the knowledge of $f_\mathrm{d}/f_\mathrm{s}$, the ratio of b-quark hadronization 
  factors that accounts for the different 
  production rate of $\Bd$ and $\Bs$ mesons. The ratio of the branching fractions of $\Bs\to\jpsi\Peta^{\prime}$ and $\Bs\to\jpsi\Peta$ 
  decays is measured to be 
\begin{equation*} 
  \frac{\BR(\Bs\to\jpsi\Peta^{\prime})}{\BR(\Bs\to\jpsi\Peta)} = 0.90\pm0.09\,(\mathrm{stat})\,^{+0.06}_{-0.02}\,(\mathrm{syst}).
  \end{equation*} 
\end{abstract}

\vspace*{0.75cm}

\begin{center}
 Submitted to Nucl. Phys. B
\end{center}
\vspace{\fill}

\end{titlepage}


\newpage
\setcounter{page}{2}
\mbox{~}
\newpage

\input{LHCb_authorlist.tex}

\cleardoublepage

%% file: LHCb_authorlist.tex
\centerline{\large\bf LHCb collaboration}
\begin{flushleft}
\small
R.~Aaij$^{38}$, 
C.~Abellan~Beteta$^{33,n}$, 
A.~Adametz$^{11}$, 
B.~Adeva$^{34}$, 
M.~Adinolfi$^{43}$, 
C.~Adrover$^{6}$, 
A.~Affolder$^{49}$, 
Z.~Ajaltouni$^{5}$, 
J.~Albrecht$^{35}$, 
F.~Alessio$^{35}$
% ...

{\footnotesize \it
% ...
$ ^{5}$Clermont Universit\'{e}, Universit\'{e} Blaise Pascal, CNRS/IN2P3, LPC, Clermont-Ferrand, France\\
$ ^{6}$CPPM, Aix-Marseille Universit\'{e}, CNRS/IN2P3, Marseille, France\\
% ...
$ ^{11}$Physikalisches Institut, Ruprecht-Karls-Universit\"{a}t Heidelberg, Heidelberg, Germany\\
% ...
$ ^{33}$Universitat de Barcelona, Barcelona, Spain\\
$ ^{34}$Universidad de Santiago de Compostela, Santiago de Compostela, Spain\\
$ ^{35}$European Organization for Nuclear Research (CERN), Geneva, Switzerland\\
% ...
$ ^{38}$Nikhef National Institute for Subatomic Physics, Amsterdam, The Netherlands\\
% ...
$ ^{43}$H.H. Wills Physics Laboratory, University of Bristol, Bristol, United Kingdom\\
% ...
$ ^{49}$Oliver Lodge Laboratory, University of Liverpool, Liverpool, United Kingdom\\
% ...
$ ^{n}$LIFAELS, La Salle, Universitat Ramon Llull, Barcelona, Spain\\
% ...
}
\end{flushleft}

%% file: intro_combined.tex
\section{Introduction}

\begin{figure}[htb]
  \setlength{\unitlength}{1mm}
  \centering
  \begin{picture}(150,45)
    \put(0,0){
      \includegraphics*[width=75mm,height=45mm,%
      ]{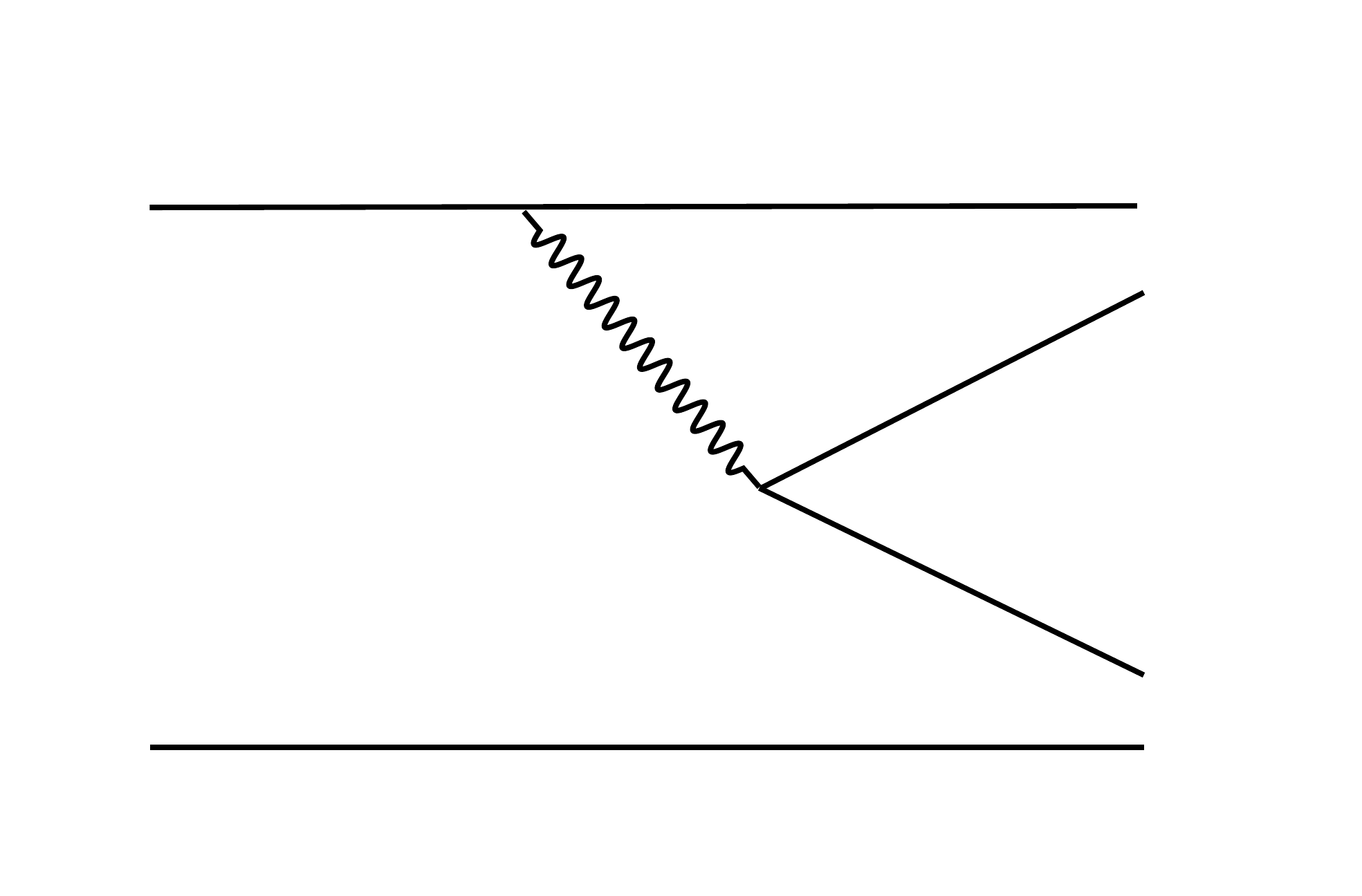}
    }
    \put(75,0){
      \includegraphics*[width=75mm,height=45mm,%
      ]{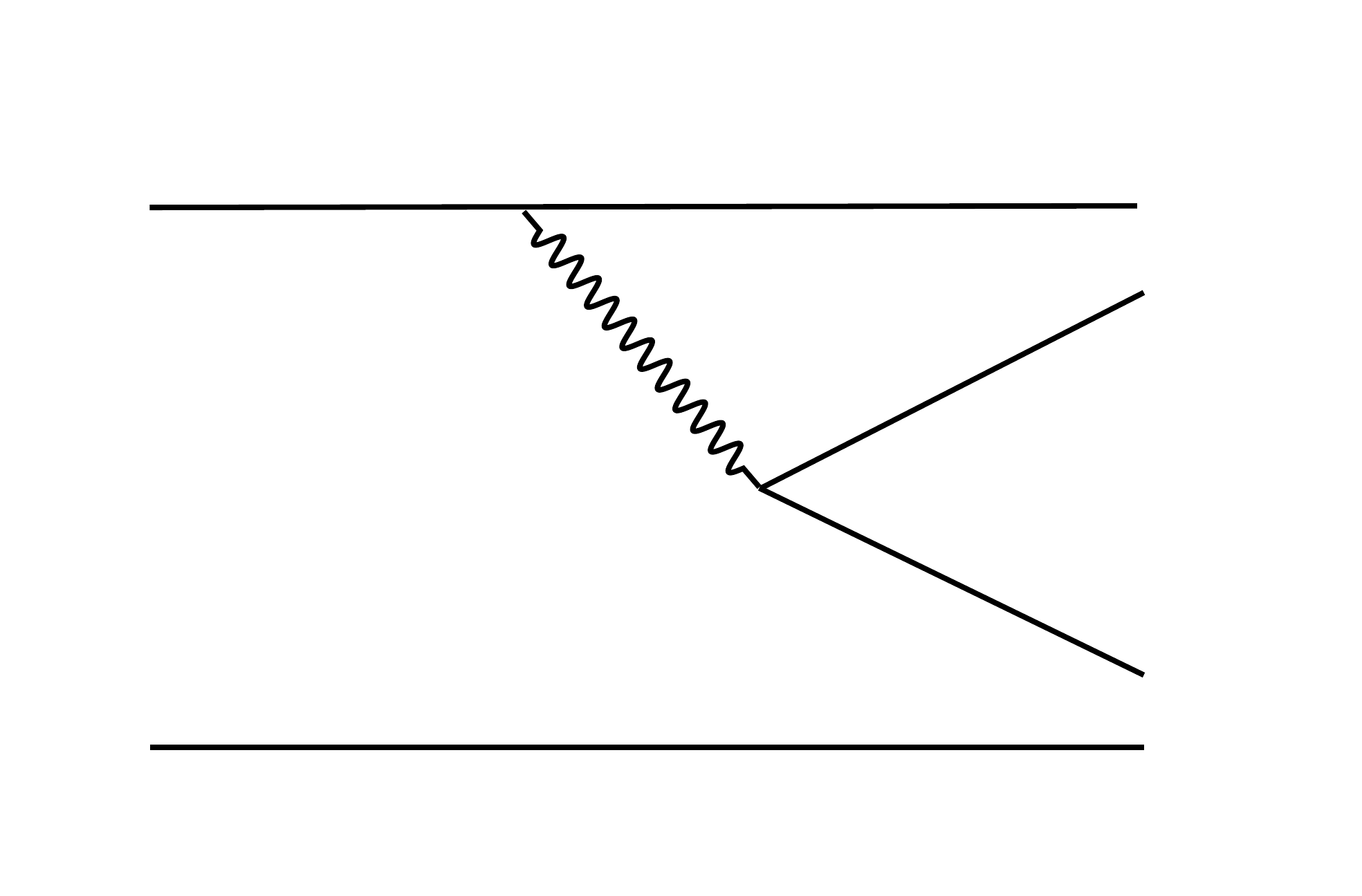}
    }
    \put(5,20)  { $\Bd$ }
    \put(10,3)  { $\mathrm{d}$ }
    \put(10,37) { $\mathrm{\bar{b}}$ }
    \put(37,27) { $\mathrm{W^-}$ }
    \put(60,3)  { $\mathrm{d}$ }
    \put(60,37) { $\mathrm{\bar{c}}$ }
    \put(60,13) { $\mathrm{\bar{d}}$ }
    \put(60,25) { $\mathrm{c}$ }
    \put(65,32) { $\jpsi$ }
    \put(65,9)  { $\Prho^0,\Pomega$ }
    \put(80,20)  { $\Bs$ }
    \put(85,3)  { $\mathrm{s}$ }
    \put(85,37) { $\mathrm{\bar{b}}$ }
    \put(112,27) { $\mathrm{W^-}$ }
    \put(135,3)  { $\mathrm{s}$ }
    \put(135,37) { $\mathrm{\bar{c}}$ }
    \put(135,13) { $\mathrm{\bar{s}}$ }
    \put(135,25) { $\mathrm{c}$ }
    \put(140,32) { $\jpsi$ }
    \put(140,9)  { $\Peta,\Peta^{\prime}$ }
  \end{picture}
  \caption {
   Examples of the dominant diagrams for the $\mathrm{B}^0_{\mathrm{(s)}}\to \jpsi \mathrm{X}^0$ decays 
   (where $\mathrm{X}^0=\Peta,\,\Peta^{\prime},\,\Pomega$ or $\Prho^0$).
  }
  \label{fig:diag}
\end{figure}

Decays of B mesons into a $\jpsi$ and a light 
meson are dominated by color-suppressed 
tree diagrams involving $\bar{\mathrm{b}}\to \bar{\mathrm{c}}\mathrm{c}\bar{\mathrm{s}}$
and $\bar{\mathrm{b}}\to \bar{\mathrm{c}}\mathrm{c}\bar{\mathrm{d}}$ transitions 
(see Fig.~\ref{fig:diag}). Contributions from other diagrams 
are expected to be small~\cite{bib:Fleischer}. Measurements of the branching
fractions of these decays can help to shed light on hadronic interactions. The decay
$\Bd\to\jpsi\Pomega$ has not been observed previously.
The CLEO collaboration has set the most restrictive upper limit to date of
$\BR(\Bd~\to~\jpsi\Pomega) \xspace < \xspace 2.7 \xspace \times \xspace 10^{-4}$ at 90\%~confidence level~\cite{Bishai:1995yj}.

The $\Bs\to\jpsi\Peta^{(\prime)}$ decays were observed by the Belle collaboration~\cite{Adachi:2009usa}
with branching fractions 
$\BR(\Bs\to\jpsi\Peta) = (5.10\pm0.50\pm0.25\,^{+1.14}_{-0.79})\times10^{-4}$ and
$\BR(\Bs\to\jpsi\Peta^{\prime}) = (3.71\pm0.61\pm0.18\,^{+0.83}_{-0.57})\times10^{-4}$,
where the first uncertainty is statistical, the second is systematic and the third one
is due to an uncertainty of the number of produced $\Bs{\mathrm{\bar B^0_s}}$ pairs. 
Since both final states are \CP eigenstates, time-dependent \CP violation
studies and access to the $\Bs-{\mathrm{\bar B^0_s}}$ mixing phase
$\Pphi_\mathrm{s}$ will be possible in the future~\cite{Jung:2012}. 
The theoretical prediction 
for these branching fractions and their ratio relies
on knowledge of the $\Peta-\Peta^{\prime}$ mixing phase $\Pphi_\mathrm{P}$.
Taking $\Pphi_\mathrm{P} = (41.4\pm0.5)^{\circ}$~\cite{bib:Kloe} and ignoring a possible 
gluonic component and corrections due to form factors, the ratio becomes
\begin{equation*}
\frac{\BR(\Bs\to \jpsi\Peta^{\prime})}{\BR(\Bs\to \jpsi\Peta)}\times
\frac{\mathcal{F}_s^{\Peta}}{\mathcal{F}_s^{\Peta^{\prime}}} = 
\frac{1}{\tan^2\Pphi_\mathrm{P}} = 1.28\,^{+0.10}_{-0.08}.
\end{equation*}
Here $\mathcal{F}_s^{\Peta^{(\prime)}}$ is
the phase space factor of the $\Bs~\to~\jpsi~\Peta^{(\prime)}$ decay and the uncertainty
is due to the inaccuracy in the knowledge of the mixing phase. As discussed in
Ref.~\cite{bib:Fleischer}, a precise measurement of this ratio tests $SU(3)$
flavour symmetry. In addition,
in combination with other measurements, the fraction of the  gluonic component
in the $\Peta^{\prime}$ meson can eventually be estimated~\cite{gluonic}.

 The analysis presented here 
is based on a data sample corresponding to an integrated luminosity of 1.0 fb$^{-1}$
collected by the LHCb detector in 2011 in pp collisions at a centre-of-mass 
energy of $\sqrt{s} = 7$~TeV.
The branching fractions of these decays are measured relative to
 $\BR(\Bd\to\jpsi\Prho^0)$ and the ratio 
$\dfrac{\BR(\Bs\to\jpsi\Peta^{\prime})}{\BR(\Bs\to\jpsi\Peta)}$
is determined.


%% file: detector_combined.tex
\section{LHCb detector}
\label{sec:det}

The \lhcb detector~\cite{Alves:2008zz} is a single-arm forward
spectrometer covering the pseudorapidity range $2<\eta <5$, designed
for the study of \bquark- and \cquark-hadrons. The
detector includes a high precision tracking system consisting of a
silicon-strip vertex detector surrounding the pp interaction region,
a large-area silicon-strip detector located upstream of a dipole
magnet with a bending power of about $4{\rm\,Tm}$, and three stations
of silicon-strip detectors and straw drift tubes placed
downstream. The combined tracking system has a momentum resolution
$\Delta p/p$ that varies from 0.4\% at 5\gevc to 0.6\% at 100\gevc,
and an impact parameter resolution of 20\mum for tracks with high
transverse momentum ($\pt$). 
Charged hadrons are identified using two ring-imaging Cherenkov (RICH) detectors. 
Photon, electron and hadron candidates are identified by a calorimeter system consisting
of scintillating-pad and pre-shower detectors, and electromagnetic and hadron calorimeters. 
Muons are identified by a system composed of alternating layers of iron and multiwire
proportional chambers.

The trigger consists of a hardware stage, based
on information from the calorimeter and muon systems, followed by a
software stage which applies a full event reconstruction.
This analysis uses events triggered by one or two muon candidates.
In the case of one muon, the hardware level requirement was for its
\pt to be larger than 1.5 \gevc; in case of two muons the restriction
$\sqrt{\pt_1 \cdot \pt_2} > 1.3$ \gevc was applied.
At the software level, the two muons were required to have an 
invariant mass in the interval
$2.97 < \mathrm{m}_{\mumu} < 3.21$~GeV/c$^2$
and to be consistent with originating from the same vertex.
To avoid the possibility that a few events with 
high occupancy dominate
the trigger processing time,
a set of global event selection requirements
based on hit multiplicities was applied.

For the simulation, pp collisions are generated using
\pythia~6.4~\cite{Sjostrand:2006za} with a specific \lhcb
configuration~\cite{LHCb-PROC-2010-056}.  Decays of hadronic particles
are described by \evtgen~\cite{Lange:2001uf} in which final state
radiation is generated using \photos~\cite{Golonka:2005pn}. The
interaction of the generated particles with the detector and its
response are implemented using the \geant
toolkit~\cite{Allison:2006ve, *Agostinelli:2002hh} as described in
Ref.~\cite{LHCb-PROC-2011-006}. The digitized output is passed 
through a full simulation of both the hardware and software 
trigger and then reconstructed in the same way as the data.

%% file: evsel_combined.tex
\section{Data sample and common selection requirements}
\label{sec:selection}

The decays $\mathrm{B}^0_{(\mathrm{s})} \to \jpsi \mathrm{X}^0$ (where
$\mathrm{X}^0$ = \Peta,
$\Peta^{\prime}$, $\Pomega$ and \pipi) are reconstructed using the $\jpsi
\to \mumu$~decay mode. The X$^0$ candidates are reconstructed in the
$\Peta\to\gamgam$, $\Peta\to\pipipi$, $\Peta^{\prime}\to\Prho^0\g$,
$\Peta^{\prime}\to\Peta\pipi$ and $\Pomega\to\pipipi$
final states. 
Pairs of oppositely charged particles identified as muons, each having $\pt > 550~\mathrm{MeV}/c$
and originating from a common vertex, are combined to form $\jpsi\to\mumu$
candidates.
Well identified muons are selected by requiring that
the difference in logarithms of the global likelihood of the muon hypothesis, 
$\Delta\ln\mathcal{L}_{\Pmu\mathrm{h}}$,
provided by the particle 
identification detectors~\cite{LHCb-PROC-2011-008}, 
with respect to the hadron hypothesis is greater than zero. 
The fit of the common two-prong vertex is required to satisfy 
$\chi^2<20$. The vertex is deemed to be well separated 
from the reconstructed primary vertex of the pp 
interaction by requiring the decay length significance
to be greater than 3.
Finally, the invariant mass of the dimuon combination is required to be 
within $\pm40~\mathrm{MeV}/c^2$~of the nominal 
\jpsi~mass~\cite{PDG2012}.

To identify charged pions the difference between
the logarithmic likelihoods of the pion and
kaon hypotheses provided by RICH detectors, 
$\Delta\ln \mathcal{L}_{\Ppi\mathrm{K}}$, 
should be greater than zero. In the reconstruction of the
$\mathrm{B}^0_{\left(\mathrm{s}\right)}\to\jpsi\pipi$ decay this requirement
is tightened to be 
$\Delta\ln \mathcal{L}_{\Ppi\mathrm{K}}>2$ 
so as to
suppress the contamination 
from $\mathrm{B}^0_{\left(\mathrm{s}\right)}\to\jpsi\Ppi\mathrm{K}$ decays
with misidentified kaons. In addition, the pion tracks are required to have $\pt > 250 \mevc$.
A minimal value of $\Delta\Pchi^2_{\mathrm{IP}}$, 
defined as the difference between the $\Pchi^2$ of the 
primary vertex, reconstructed with and without the 
considered track,
is required to be larger than four.

Photons are selected from neutral clusters in the electromagnetic
 calorimeter with minimal transverse energy in excess of $300~\mathrm{MeV}$. 
To suppress the large combinatorial background 
from $\piz\to\gamgam$~decays, photons that can form 
part of a $\piz\to\gamgam$~candidate 
with invariant mass within $\pm25~\mathrm{MeV}/c^2$~of the 
nominal \piz~mass are not used for 
reconstruction of $\Peta\to\gamgam$~and  
$\Peta^{\prime}\to\Prho^0\g$~candidates.
 
The $\Peta\to\gamgam$\,($\piz\to\gamgam$)~candidates are reconstructed as
diphoton combinations with  invariant mass within
$\pm70\,(25)~\mathrm{MeV}/c^2$ around the nominal $\Peta\,(\piz)$~mass.
To suppress the combinatorial background to the $\Peta\to\gamgam$ decay,
 the cosine of the decay angle $\theta^*_{\Peta}$,
between the photon momentum in the $\Peta$~rest frame and the 
direction of the Lorentz boost from the laboratory frame to 
the $\Peta$~rest frame, is required to
have $\left| \cos \theta^*_{\Peta} \right| < 0.8$.

The $\Peta^{\prime}$~candidates 
are reconstructed 
as $\Peta\pipi$~and $\Prho^0\g$~combinations with 
invariant mass within
$\pm60~\mathrm{MeV}/c^2$ from the 
nominal $\Peta^{\prime}$~mass.
For the $\Peta^{\prime}\to\Prho^0\g$~case, 
the invariant mass of the \pipi~combination is required to be 
within $\pm150~\mathrm{MeV}/c^2$ of the $\Prho^0$~mass.
For $\Peta\to\pipipi$~($\Pomega\to\pipipi$)~candidates 
the invariant mass is required to be within $\pm50~\mathrm{MeV}/c^2$ 
of the nominal $\Peta\,(\Pomega)$~mass.

The $\mathrm{B}^0_{\left(\mathrm{s}\right)}$~candidates are 
formed from $\jpsi \mathrm{X}^0$~pairs
with $\pt > 3 \gevc$ for the $\mathrm{X}^0$. To improve the invariant mass 
resolution a kinematic fit~\cite{Hulsbergen:2005pu} is applied.
 In this fit, constraints are applied on the known masses~\cite{PDG2012} of 
intermediate resonances, except the wide $\Prho^0$ and $\Pomega$ states, and 
it is also required that the candidate's momentum vector points to the 
associated primary vertex. The $\chi^2$ per degree of freedom for 
this fit is required to be less than five. Finally, the decay time (c$\Ptau$) of the
$\mathrm{B}^0_{\left(\mathrm{s}\right)}$~candidates is required
 to be in excess of~$150 \mum$.

%% file: ana2_combined.tex
\section{Evidence for the $\boldsymbol{\Bd\to\jpsi\Pomega}$ decay}
\label{sec:omega}

\begin{figure}[t]
  \setlength{\unitlength}{1mm}
  \centering
  \begin{picture}(150,120)
    \put(0,0){
      \includegraphics*[width=150mm,height=120mm,%
      ]{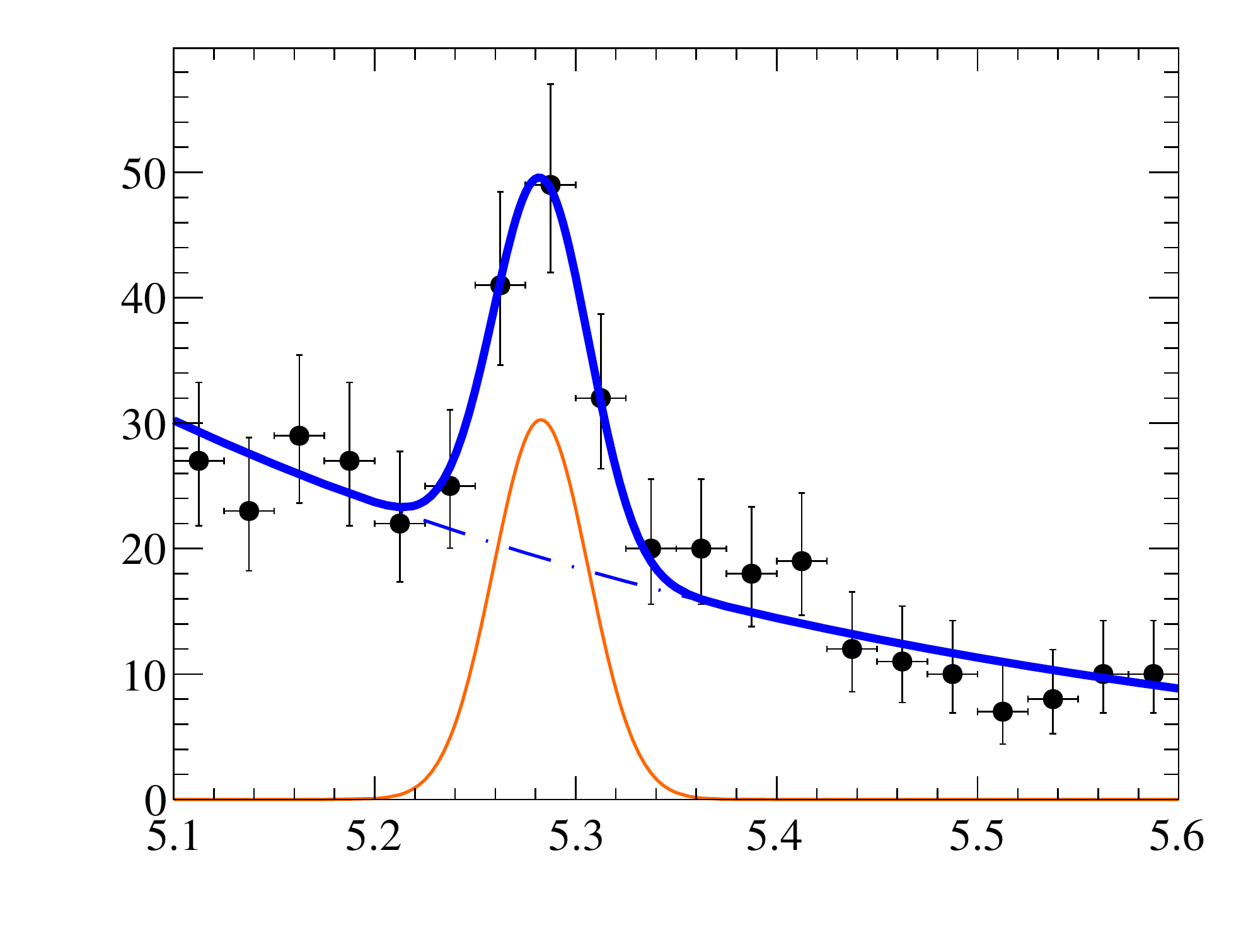}
    }
    \put(75,6)   { {\Large$\mathrm{m}$}$_{\jpsi\Pomega}$ }
    \put(123,6)  { \large $\left[ \mathrm{GeV}/c^2\right]$ }
    \put(115,100)  { \Large{LHCb} }
    \put(30,100)  { \Large{$\Bd \rightarrow \jpsi\Pomega$} }
    \put(5 ,52)  {
      \begin{sideways}%
        \Large{Candidates/$\left(25~\mathrm{MeV}/c^2 \right)$}
      \end{sideways}%
    }

  \end{picture}
  \caption {
    Invariant mass distribution for selected $\Bd\to\jpsi\Pomega$~candidates.
    The black dots correspond to the data
    distribution, the thick solid blue line is the total fit function, the
    blue dashed line shows the background contribution and the orange thin
    line is the signal component of the fit function.
  }
  \label{fig:bd_2_psi_omega}
\end{figure}
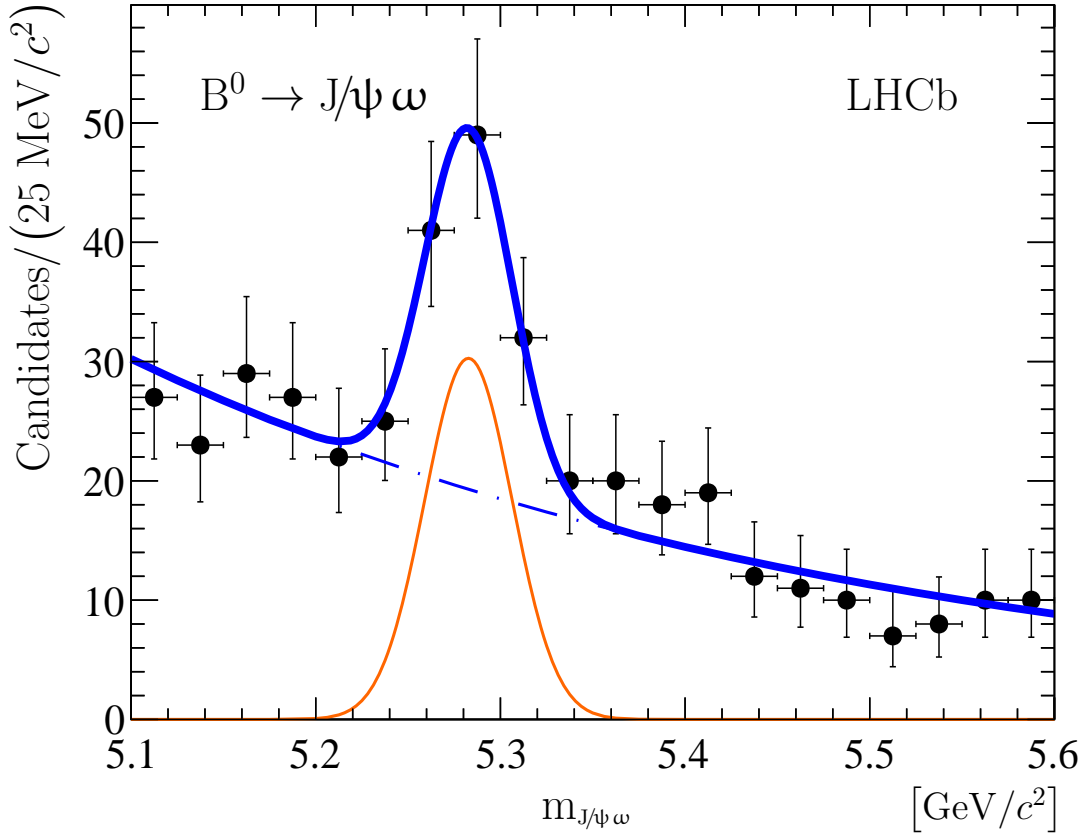

The invariant mass distribution of the selected $\jpsi\Pomega$ candidates is shown 
in Fig.~\ref{fig:bd_2_psi_omega}, where a $\Bd$ signal is visible.
To determine the signal yield, an unbinned maximum likelihood fit is performed to this distribution.
The signal is modelled by a Gaussian distribution and the
background by an exponential function. The peak position is found to be $5284\pm5$~\.
MeV/$c^2$, which is consistent with the nominal $\Bd$ mass~\cite{PDG2012}
and the resolution is in good agreement with the prediction from simulation.
The event yield is determined to be $\mathcal{Y}_{\Bd} = 72\pm15$.

The statistical significance for the observed
signal is determined as
\mbox{ $
 \mathcal{S} = \sqrt{-2\times \ln
       (  \mathcal{L}_{\mathcal{B}} / \mathcal{L}_{\mathcal{S}+\mathcal{B}}) },
 $}
where $\mathcal{L}_{\mathcal{S}+\mathcal{B}}$ and
      $\mathcal{L}_{\mathcal{B}}$
denote the likelihood of the signal plus background hypothesis and
the background hypothesis, respectively. 
The statistical significance of the 
signal is found to be 5.0 standard deviations.
Taking into account the systematic uncertainty
related to the fit function, which is 
discussed in detail in Section~\ref{sec:syst}, the significance is
4.6\Psigma; this also takes into account the freedom in the
peak position and width in the nominal fit.

To demonstrate that the signal originates from
$\Bd\to\jpsi\Pomega$~decays, the sPlot
technique~\cite{Pivk:2004ty} has been applied.
Using the $\jpsi\pipi\gamgam$ invariant mass as the discriminating variable, 
the  distributions for the invariant masses of
the intermediate resonances $\piz\to\gamgam$ and 
$\Pomega\to\pipipi$ have been obtained.
The invariant mass window for each corresponding
resonance is released and the mass constraint is removed.

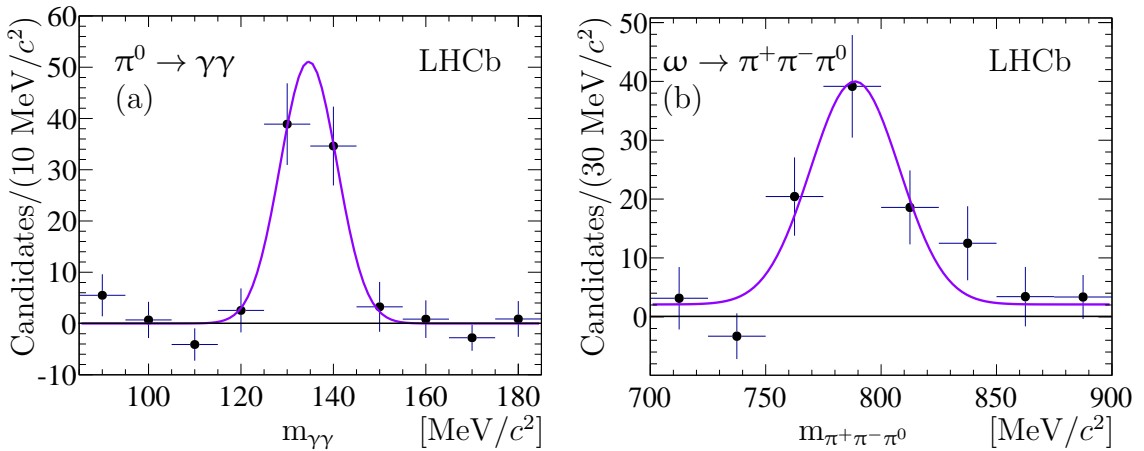
\begin{figure}[t]
  \setlength{\unitlength}{1mm}
  \centering
  \begin{picture}(150,60)
    \put(0,0){
      \includegraphics*[width=75mm,height=60mm,%
      ]{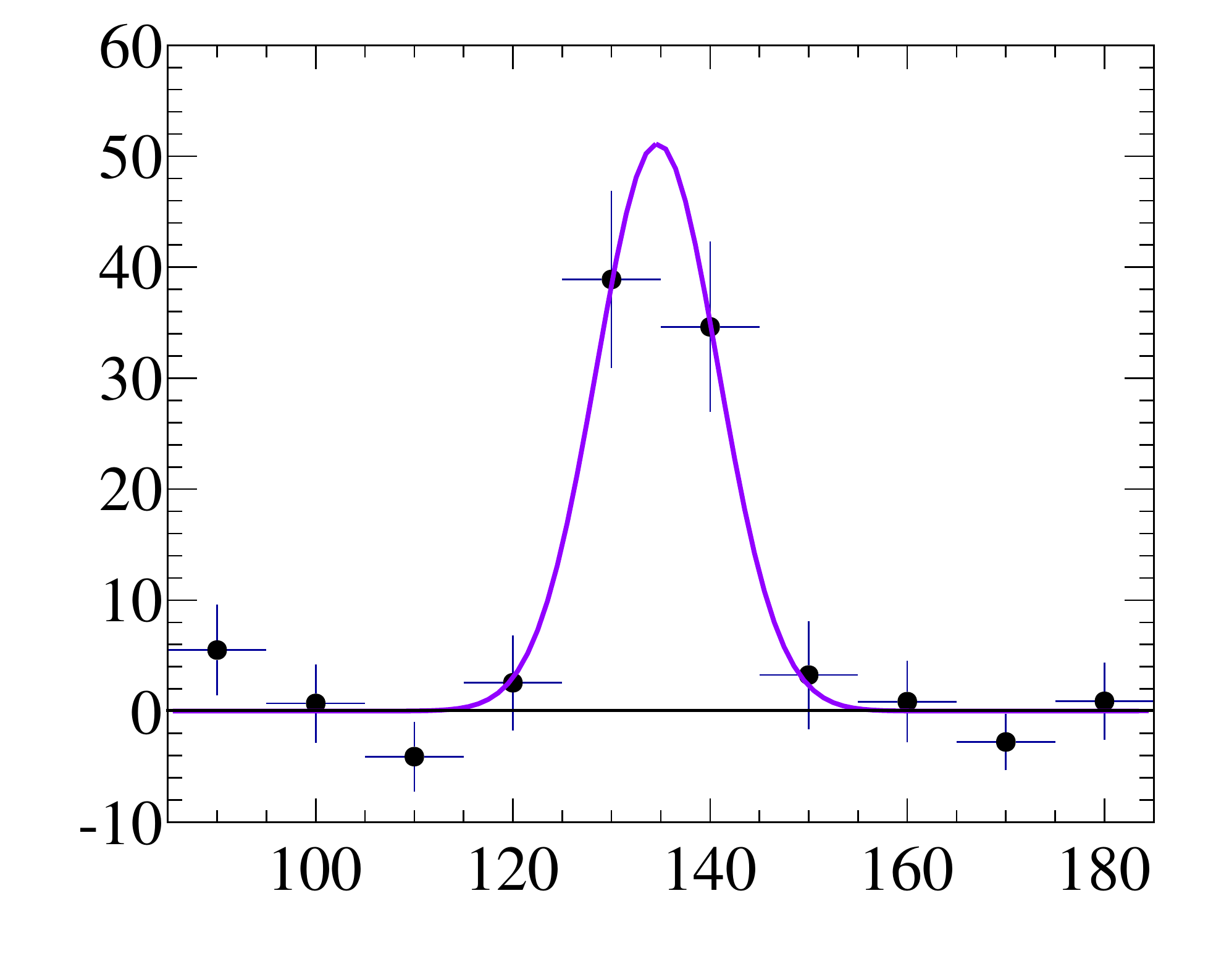}
    }
    \put(15,50) { $\piz \to \gamgam$  }
    \put(37,1)   { $\mathrm{m}_{\gamgam}$  }
    \put(55,1)  { $\left[ \mathrm{MeV}/c^2\right]$}
    \put(55,50)  { \text{LHCb} }
    \put(15,45)  { (a) }
    \put(1,12)  {
      \begin{sideways}%
        Candidates/$\left(10~\mathrm{MeV}/c^2\right)$
      \end{sideways}%
    }
    \put(75,0){
      \includegraphics*[width=75mm,height=60mm,%
      ]{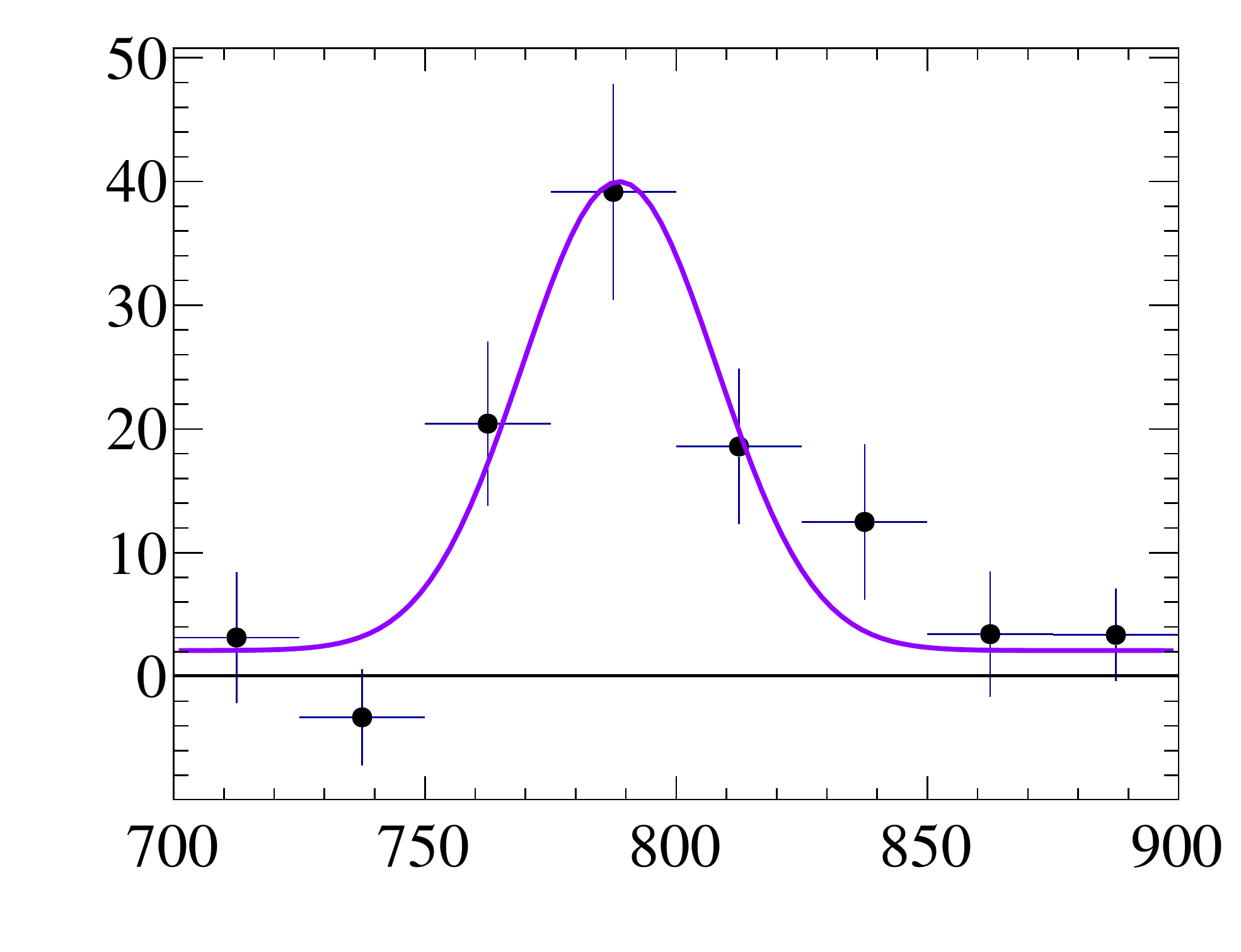}
    }
    \put(87,50) { $\Pomega \to \pipipi$  }
    \put(105,1)   { $\mathrm{m}_{\pipipi}$  }
    \put(130,1)  { $\left[ \mathrm{MeV}/c^2\right]$}
    \put(130,50)  { \text{LHCb} }
    \put(87,45)  { (b) }
    \put(76 ,12)  {
      \begin{sideways}%
        Candidates/$\left(30~\mathrm{MeV}/c^2 \right)$
      \end{sideways}%
    }
  \end{picture}
  \caption {
    Background-subtracted (a) $\gamgam$
    and (b) $\pipipi$ invariant mass distributions 
   for $\Bd\to\jpsi\pipi\gamma\gamma$ decays. In both
   distributions the line is the result of the fit
   described in the text.
  }
  \label{fig:splot_Omega}
\end{figure}

The invariant mass distributions 
for \gamgam and \pipipi from 
$\Bd\to\jpsi\Pomega$~candidates are shown in 
Fig.~\ref{fig:splot_Omega}.
Clear signals are seen for both the
           $\Pomega\to\pipipi$ and
           $\piz\to\gamgam$ decays.
The \gamgam distribution is described by a sum
of a Gaussian function and a constant. The $\Pomega\to\pipipi$ signal
is modelled by a convolution of a Gaussian and a Breit-Wigner function
with a constant background. The peak positions are in good agreement 
with the nominal \piz and $\Pomega$~masses and the yields determined
from the fits are compatible with the $\Bd\to\jpsi\Pomega$ yield. The
 nonresonant contribution in each case is found to be consistent with zero.

%% file: ana_combined.tex

\section{Decays into $\boldsymbol{\jpsi\Peta^{(\prime)}}$ final states}
\label{sec:ana}
\begin{figure}[t]
  \setlength{\unitlength}{1mm}
  \centering
  \begin{picture}(156,124)
    \put(0,62){
      \includegraphics*[width=78mm,height=62mm%
      ]{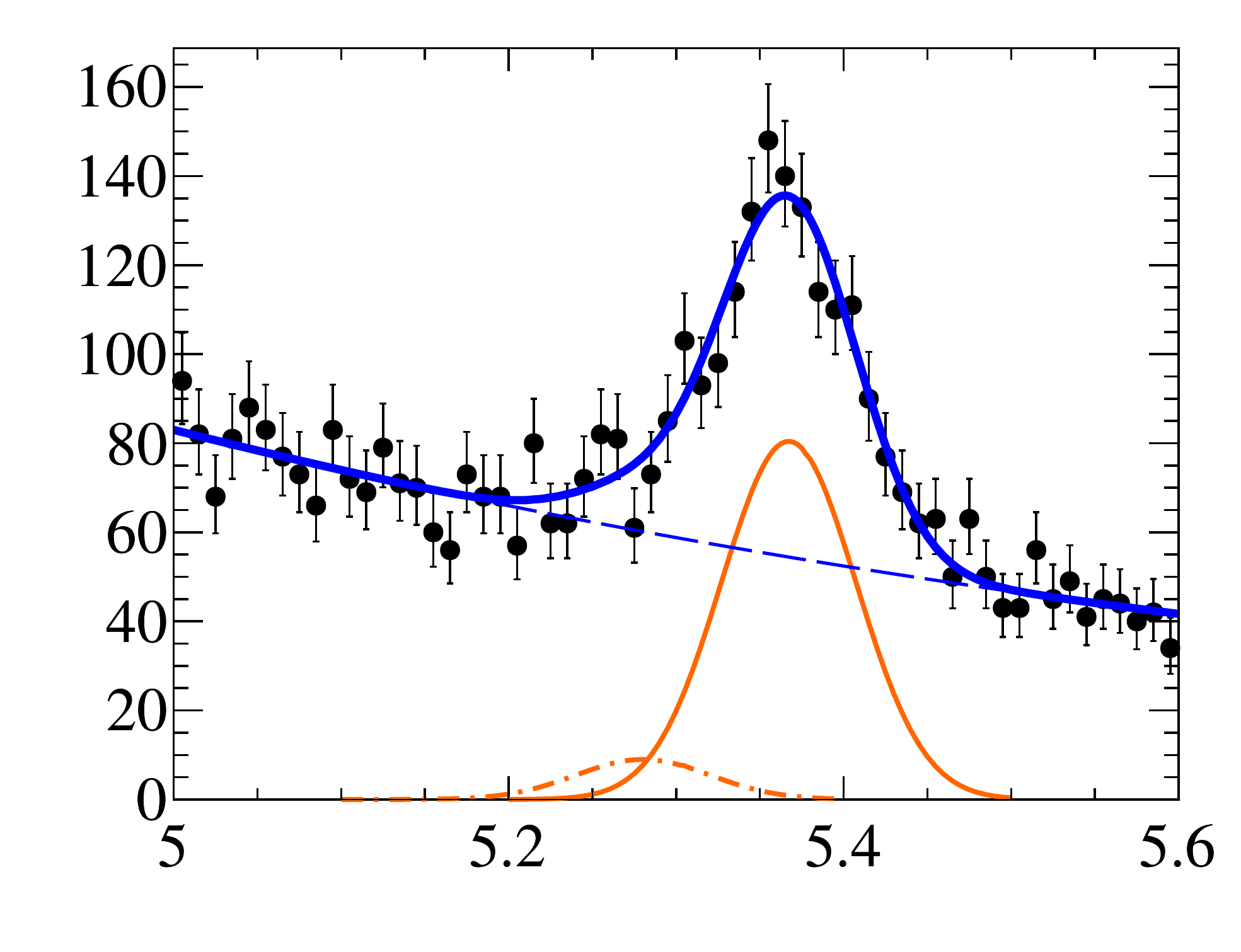}
    }
    \put(0,0){
      \includegraphics*[width=78mm,height=62mm,%
      ]{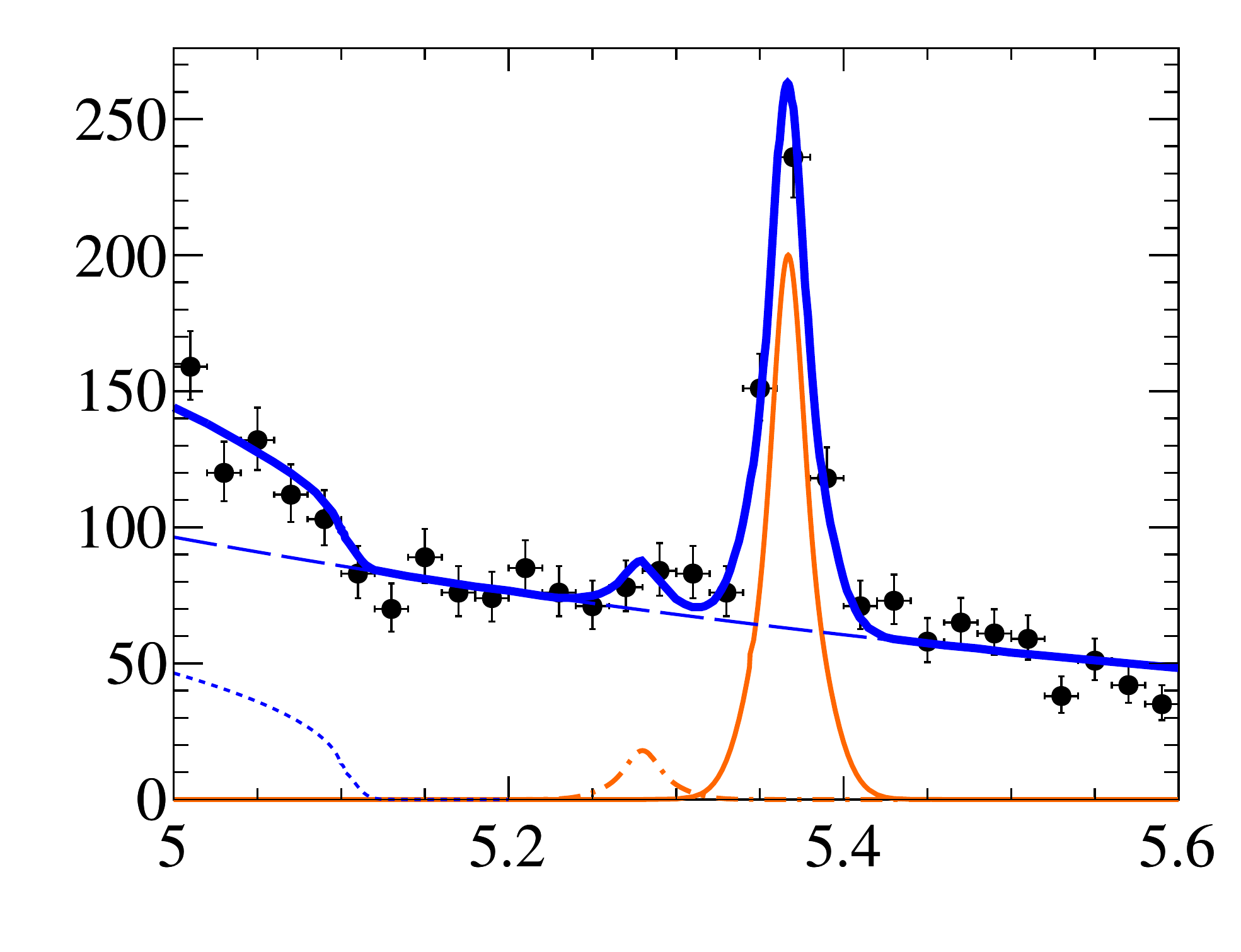}
    }
    \put(78,0){
      \includegraphics*[width=78mm,height=62mm,%
      ]{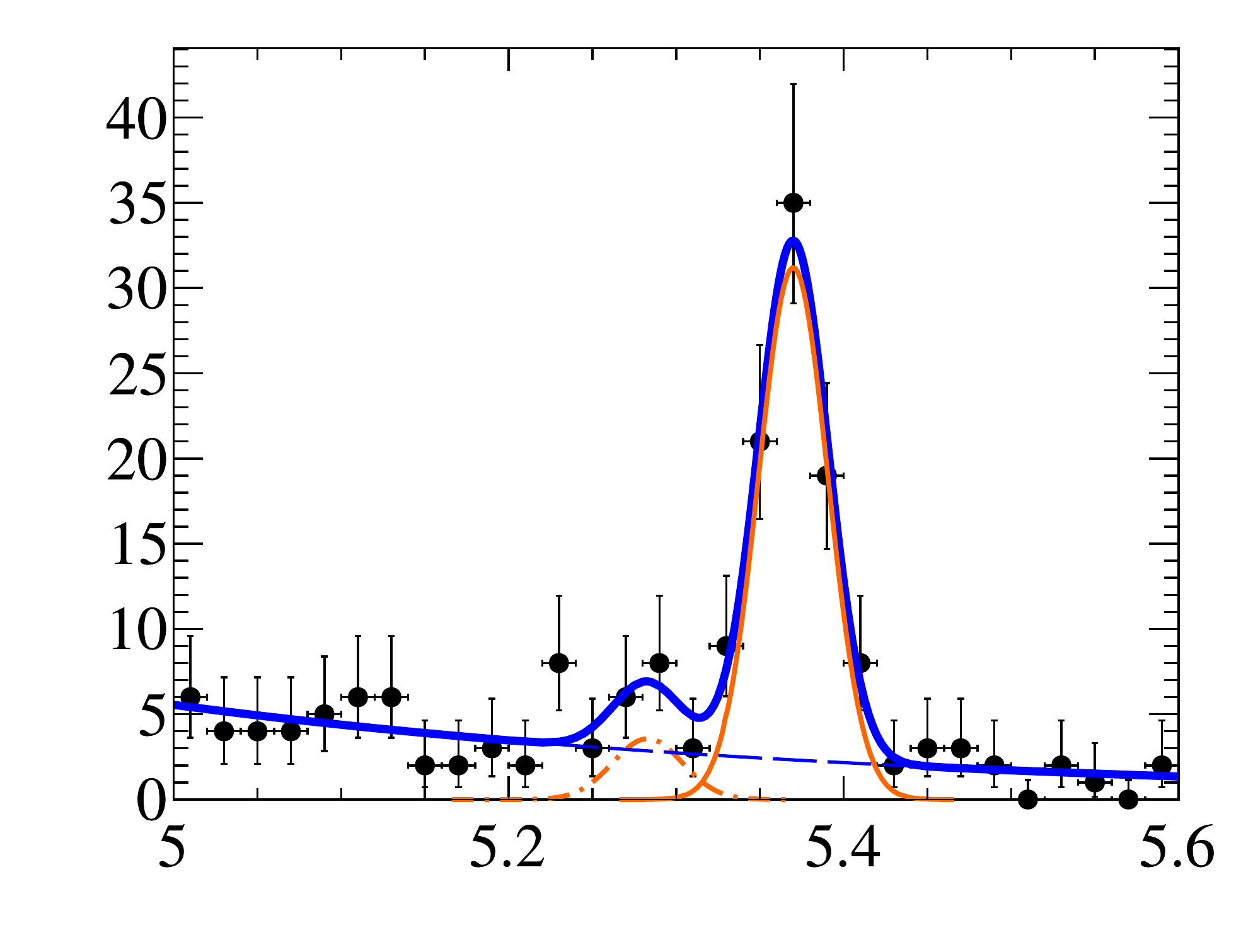}
    }
    \put(78,62){
      \includegraphics*[width=78mm,height=62mm%
      ]{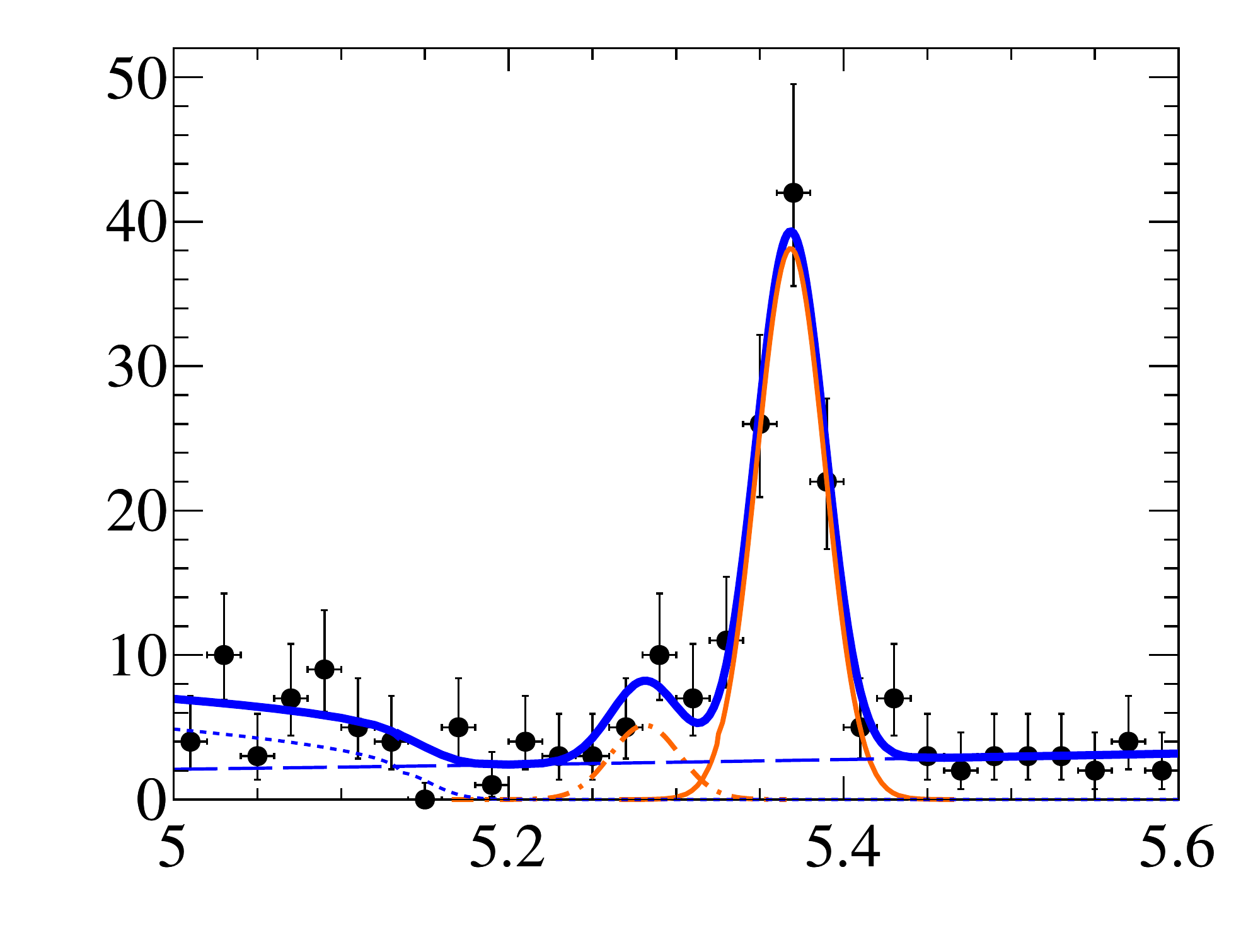}
    }
    \put(14, 115) {(a)}
    \put(19 ,115) { \small $\Bs\to\jpsi\Peta$} 
    \put(19 ,110) { \small $(\Peta\to\gamgam$)} 
    \put(61, 115) {LHCb}
    \put(93, 115) {(b)}
    \put(99,115) {\small $\Bs\to\jpsi\Peta$} 
    \put(99,110) {\small $(\Peta\to3\Ppi)$} 
    \put(139, 115) {LHCb}
    \put(15, 52) {(c)}
    \put(21 ,52) {\small $\Bs\to\jpsi\Peta^{\prime}$} 
    \put(21 ,47) {\small $(\Peta^{\prime}\to\Prho^0\g)$} 
    \put(61, 52) {LHCb}
    \put(93, 52) {(d)}
    \put(98,52) { \small $\Bs\to\jpsi\Peta^{\prime}$} 
    \put(98,47) { \small $(\Peta^{\prime}\to\Ppi\Ppi\Peta)$} 
    \put(139, 52) {LHCb}
    \put( 37,63)  { $\mathrm{m}_{\jpsi\Peta}$ }
    \put(117,63)  { $\mathrm{m}_{\jpsi\Peta}$ }
    \put( 59,63)  { $\left[ \mathrm{GeV}/c^2\right]$}
    \put(136,63)  { $\left[ \mathrm{GeV}/c^2\right]$}
    \put( 37,1)  { $\mathrm{m}_{\jpsi\Peta^{\prime}}$ }
    \put(117,1)  { $\mathrm{m}_{\jpsi\Peta^{\prime}}$ }
    \put( 58,1)  { $\left[ \mathrm{GeV}/c^2\right]$}
    \put(136,1)  { $\left[ \mathrm{GeV}/c^2\right]$}
    \put(-1 ,16)  { \small
      \begin{sideways}%
        Candidates/$\left(20~\mathrm{MeV}/c^2 \right)$
      \end{sideways}%
    }
    \put(80 ,16)  { \small
      \begin{sideways}%
        Candidates/$\left(20~\mathrm{MeV}/c^2\right)$
      \end{sideways}%
    }
    \put(-1 ,78)  { \small
      \begin{sideways}%
        Candidates/$\left(10~\mathrm{MeV}/c^2\right)$
      \end{sideways}%
    }
    \put(80 ,78)  { \small
      \begin{sideways}%
        Candidates/$\left(20~\mathrm{MeV}/c^2 \right)$
      \end{sideways}%
    }
  \end{picture}
  \caption {
    Invariant mass distributions for selected $\Bs\to\jpsi\Peta^{\left(\prime\right)}$~candidates:
    (a)~\mbox{$\Bs\to\jpsi\Peta\left(\Peta\to\gamgam\right)$},
    (b)~$\Bs\to\jpsi\Peta\left(\Peta\to\pipipi\right)$,
    (c)~$\Bs\to\jpsi\Peta^{\prime}\left(\Peta^{\prime}\to\Prho^0\g\right)$ and 
    (d)~$\Bs\to\jpsi\Peta^{\prime}\left(\Peta^{\prime}\to\pipi{}\Peta\right)$. In all distributions
    the black dots show the data. The thin solid orange lines show the signal $\Bs$
    contributions and the orange dot-dashed lines correspond to the $\Bd$ contributions.
    The blue dashed lines show the combinatorial background contributions and the dotted
    blue lines show the partially reconstructed background components. The total fit
    functions are drawn as solid blue lines.
    The results of the fit are described in the text.
  }
  \label{fig:bs_2_psi_eta_etap}
\end{figure}

The invariant mass spectra for 
$\Bs\to\jpsi\Peta^{\left(\prime\right)}$~candidates 
are shown in Fig.~\ref{fig:bs_2_psi_eta_etap}, where signals are visible.
To determine
the signal yields, unbinned maximum likelihood fits are performed. For
all modes apart from $\jpsi\Peta^{\prime}\left(\Peta^{\prime}\to\Prho^0\g\right)$,
 the $\Bs$ signal is modelled by a single Gaussian function. In all cases there 
is a possible corresponding
$\Bd$ signal, which is included in the fit model as
an additional Gaussian component. The difference of the means of the two Gaussians 
is fixed to the known difference between the $\Bs$ and the $\Bd$ masses~\cite{Aaij2012bmass}. 
Simulation studies for the $\jpsi\Peta^{\prime}\left(\Peta^{\prime}\to\Prho^0\g\right)$ 
mode indicate that in this case a
double Gaussian resolution model is more appropriate. 
The mean values of the two Gaussian functions are required
to be the same, and the ratio of their
resolutions and the fraction of the event yield
carried by each of the Gaussian functions are fixed at
the values obtained from simulation.

The combinatorial background is modelled by an exponential function.
 In addition, a component is added to describe the contribution from partially 
reconstructed $\mathrm{B}$ decays. It is described 
with the phase space function for two 
particles in a three body decay under the hypothesis
of $\mathrm{B}\to\jpsi\Peta^{(\prime)}\mathrm{X}$ decay,
where $\mathrm{X}$ can be either a kaon or a pion, which escapes
 detection. The phase space function is convolved with a resolution 
factor, which is fixed at the value of the signal resolution.

The fit results are summarized in
Table~\ref{tab:ana_signal}.
In all cases the position of the signal peak is consistent 
with the nominal $\Bs$ mass~\cite{PDG2012}
and the resolutions agree with the expectations from simulation.
The statistical significances of all the $\Bs$ decays exceed 7$\Psigma$.
\begin{table}[t]
  \centering
  \caption{
   Signal yields, $\mathcal{Y}_{\Bs}$,
   the fitted \Bs~mass,
   $\mathrm{m}_{\Bs}$ and 
   mass resolutions, 
   $\sigma_{\Bs}$ for the
   $\Bs\to\jpsi{}\Peta^{\left(\prime\right)}$ decays. 
  } \label{tab:ana_signal}
  \vspace*{3mm}
  \begin{tabular*}{0.90\textwidth}{@{\hspace{5mm}}l@{\extracolsep{\fill}}ccc@{\hspace{5mm}}}
    \multirow{2}{*}{Mode} 
    &  \multirow{2}{*}{$ \mathcal{Y}_{\Bs}$} 
    &  $ \mathrm{m}_{\Bs}$ 
    &  $ \sigma_{\Bs}$ 
    \\
    &  
    &  {\small{$ \left[\mathrm{MeV}/c^2\right]$}} 
    &  {\small{$ \left[\mathrm{MeV}/c^2\right]$}}
    \\
    \hline
    $\Bs \rightarrow \jpsi \Peta(\Peta\to\gamgam)$
    & \multirow{1}*{$810\pm65$}
    & \multirow{1}*{$5367.2\pm3.5$}
    & \multirow{1}*{$40.1\pm3.6$}
    \\
    $\Bs \rightarrow \jpsi \Peta(\Peta\to\pipipi)$
    & \multirow{1}*{\;\;$94\pm11$}
    & \multirow{1}*{$5368.4\pm2.6$}
    & \multirow{1}*{$20.3\pm2.3$}
    \\
    $\Bs \rightarrow \jpsi \Peta^{\prime}(\Peta^{\prime}\to\Prho^0{}\g)$
    & \multirow{1}*{$336\pm30$}
    & \multirow{1}*{$5367.0\pm1.1$}
    & \multirow{1}*{\;$8.0\:\pm1.1$}
    \\
    $\Bs \rightarrow \jpsi \Peta^{\prime}(\Peta^{\prime}\to\pipi{}\Peta)$
    & \multirow{1}*{\;\;$79\pm10$}
    & \multirow{1}*{$5369.0\pm2.8$}
    & \multirow{1}*{$20.7\pm2.3$}
    \\
  \end{tabular*}   
\end{table}

To test the resonance structure of the $\Bs\to\jpsi\Peta^{(\prime)}$ decays,
 the sPlot technique is used. For 
the $\piz$, $\Peta$ and $\Peta^{\prime}$~candidates the background-subtracted
invariant mass distributions are studied. The restrictions 
on the invariant mass for the corresponding
resonance are released and the mass constraints 
(if any) removed.
The background-subtracted distributions
are then fitted with the sum of a Gaussian 
function and a constant component for the resonant 
and nonresonant components respectively.
 In the fit of the
 dipion~invariant mass for the $\Peta^{\prime}\to\pipi\g$~decay a
modified relativistic Breit-Wigner function is used as
 the signal component~\cite{Jackson:1964zd,Selleri:1962}.

\begin{figure}[t]
  \setlength{\unitlength}{1mm}
  \centering
  \begin{picture}(150,170)
    \put(0,120){
      \includegraphics*[width=75mm,height = 60mm%
      ]{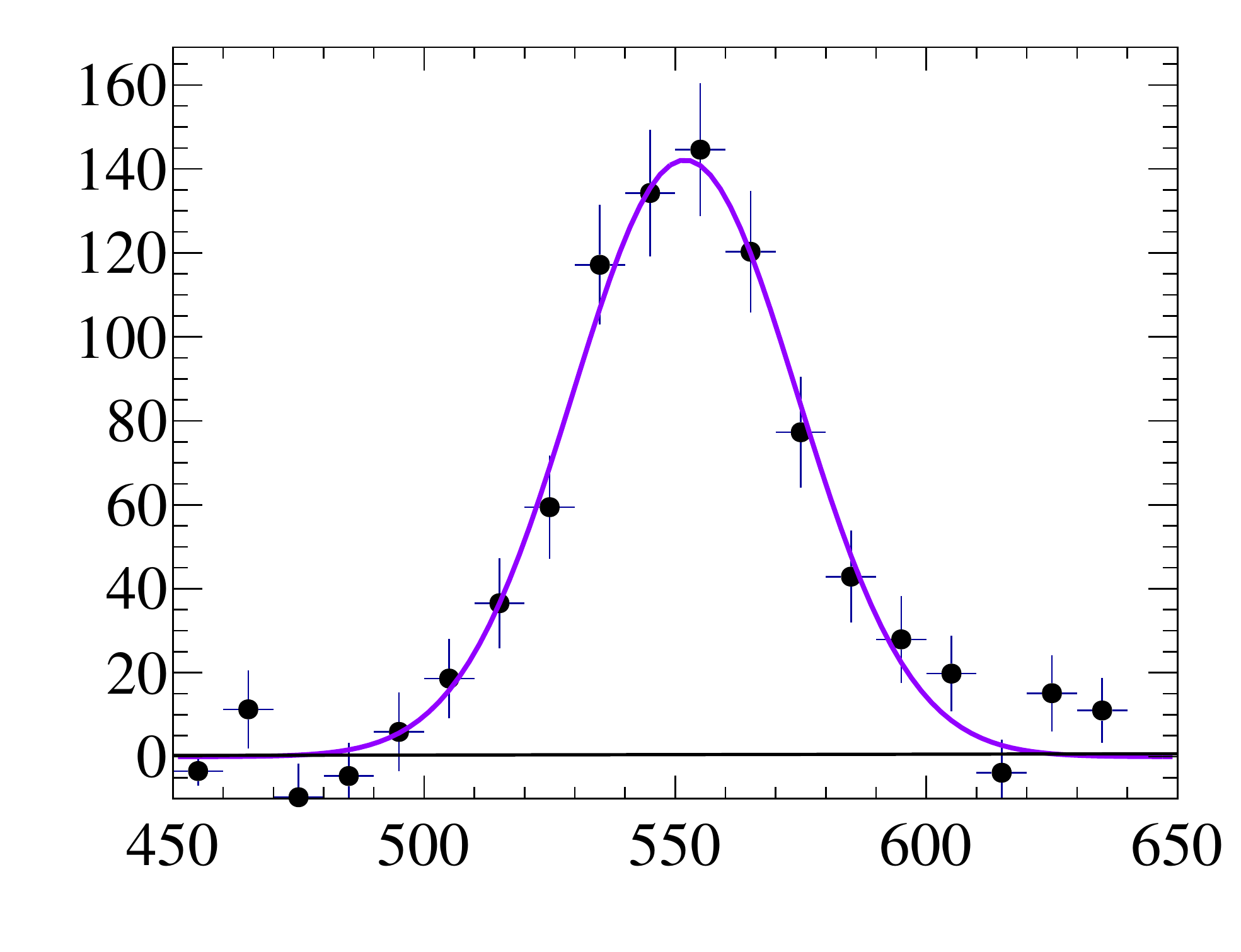}
    }
    \put(75,120){
      \includegraphics*[width=75mm,height = 60mm%
      ]{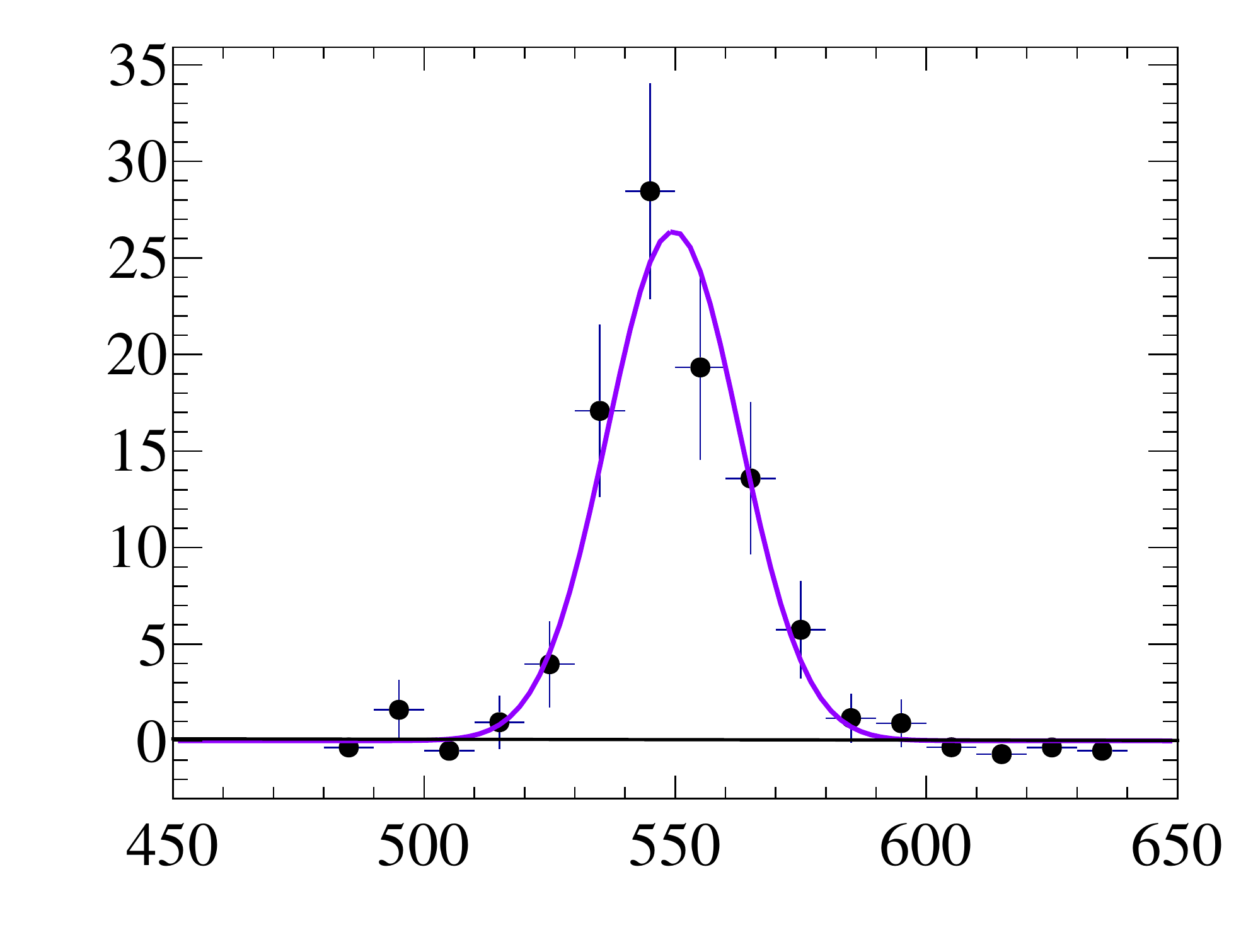}
    }
    \put(0,60){
      \includegraphics*[width=75mm,height = 60mm%
      ]{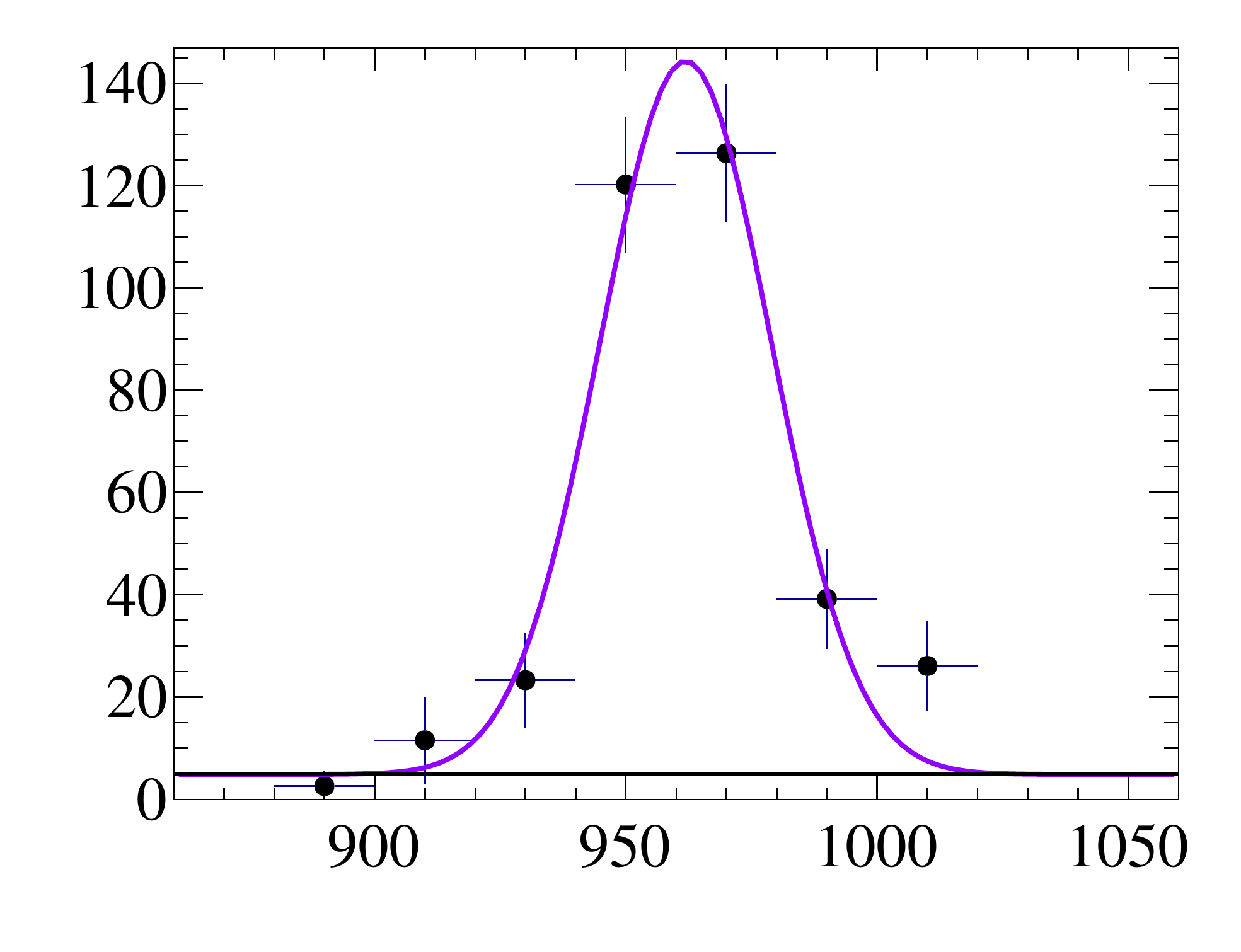}
    }
    \put(75,60){
      \includegraphics*[width=75mm,height = 60mm%
      ]{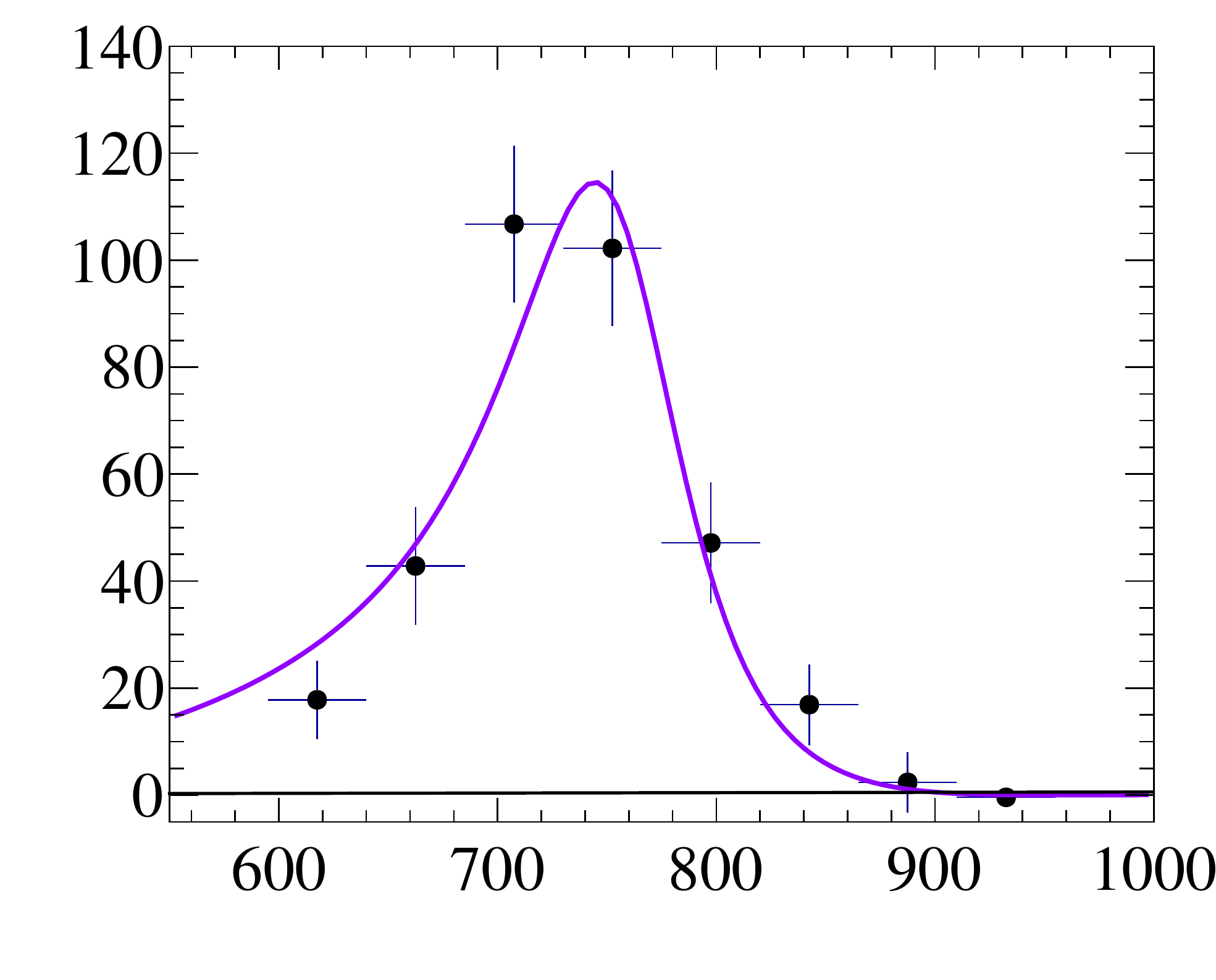}
    }
    \put(0,0){
      \includegraphics*[width=75mm,height = 60mm%
      ]{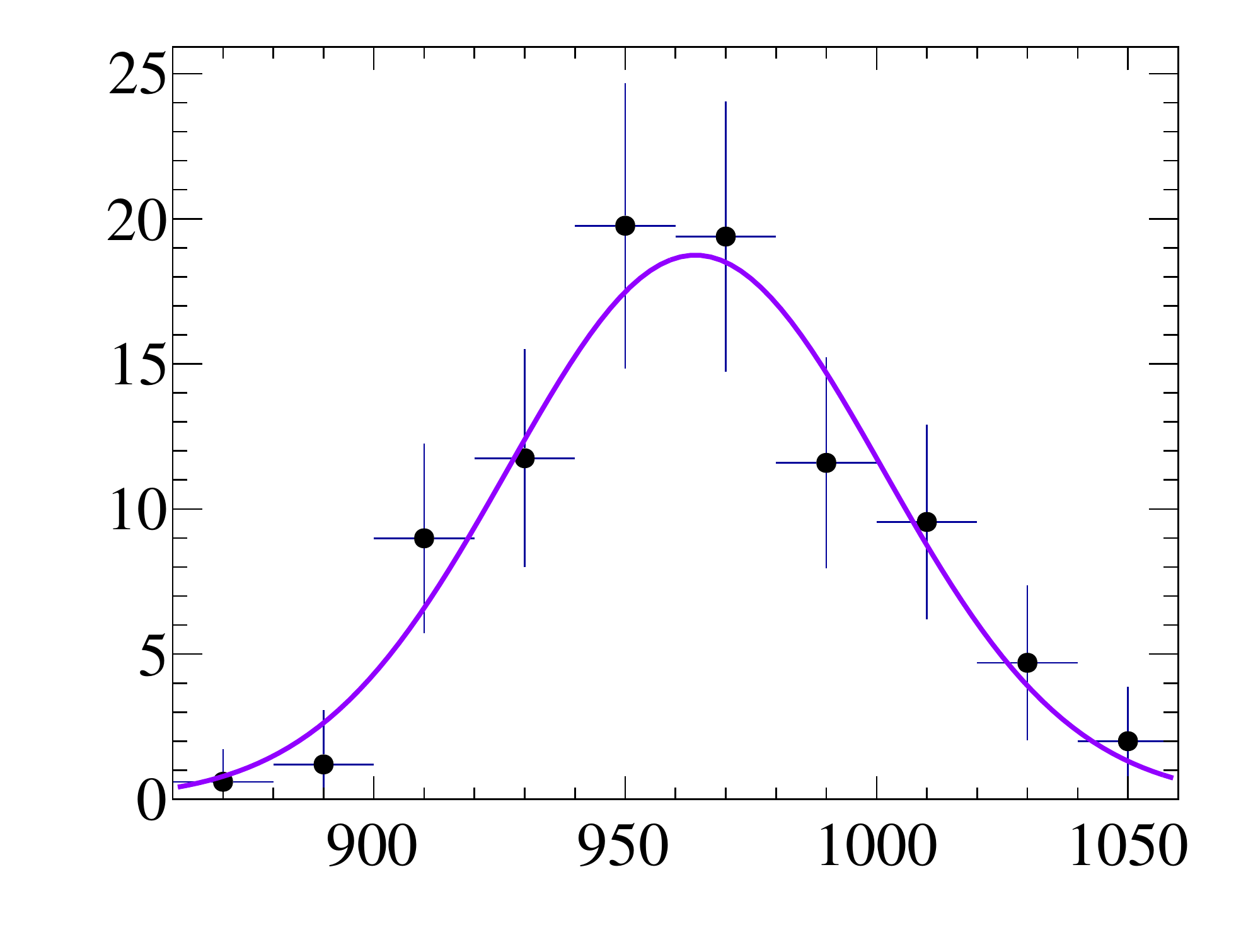}
    }
    \put(75,0){
      \includegraphics*[width=75mm,height = 60mm%
      ]{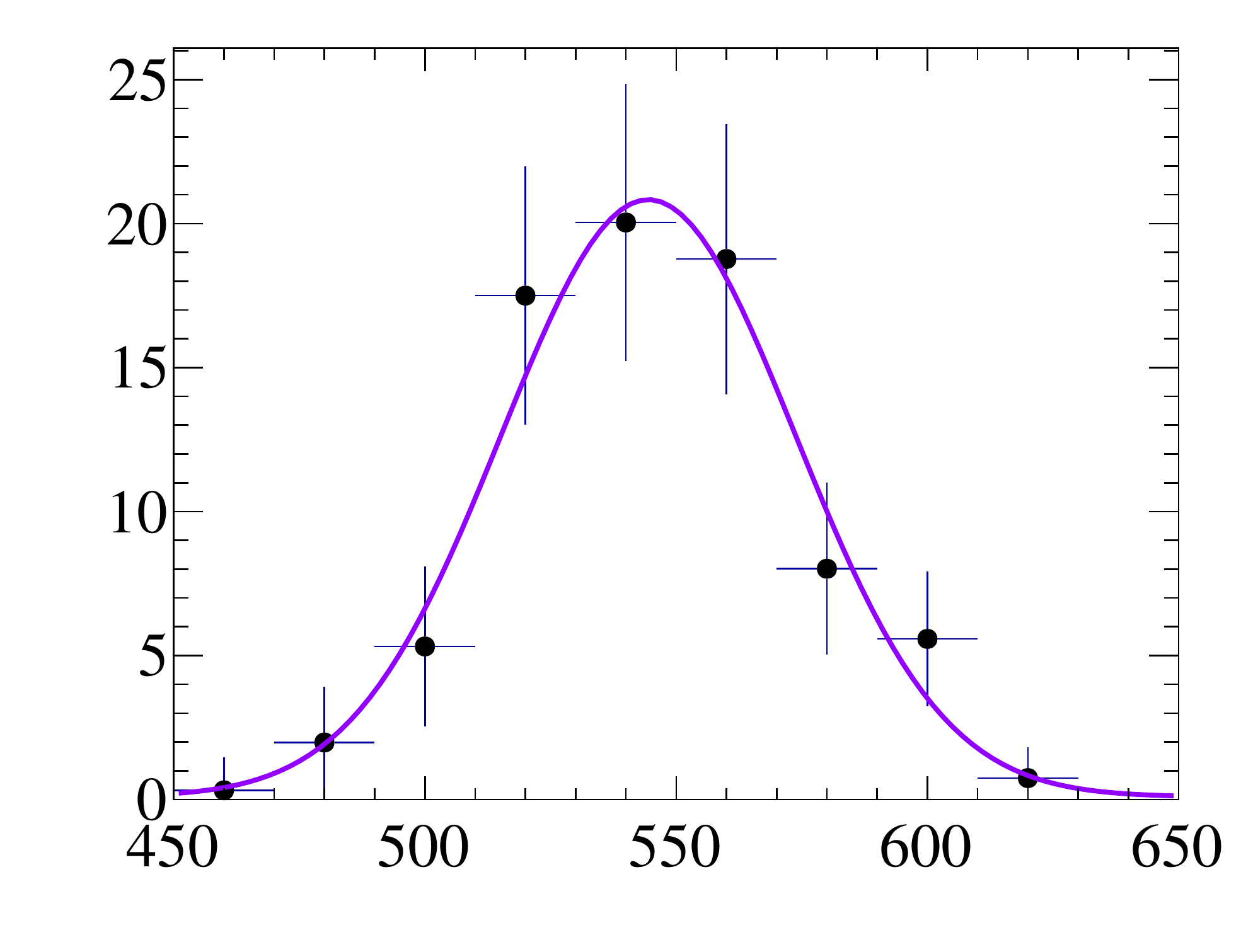}
    }
    \put(15,165) {\small (a)  } 
    \put(15,170) {\small $\Peta \to \g\g$  } 
    \put(55,170) { LHCb  } 
    \put(37,121) {\small $\mathrm{m}_{\g\g}$ }
    \put(57,121)  {\small $\left[ \mathrm{MeV}/c^2\right]$}
    \put(0,133)  { 
      \begin{sideways}%
        {\small Candidates/$\left(10~\mathrm{MeV}/c^2\right)$}
      \end{sideways}%
    }
    \put(90,165) {\small (b)  } 
    \put(90,170) {\small $\Peta \to \pipipi$  } 
    \put(130,170) { LHCb  } 
    \put(112,121) {\small $\mathrm{m}_{\pipipi}$ }
    \put(132,121)  {\small $\left[ \mathrm{MeV}/c^2\right]$}
    \put(76,133)  { 
      \begin{sideways}%
        {\small Candidates/$\left(10~\mathrm{MeV}/c^2\right)$}
      \end{sideways}%
    }
    \put(15,105) {\small (c)  } 
    \put(15,110) {\small $\Peta^{\prime} \to \Prho^0\g$  } 
    \put(55,110) { LHCb  } 
    \put(37,61) {\small $\mathrm{m}_{\Prho^0\g}$  }
    \put(57,61)  {\small $\left[ \mathrm{MeV}/c^2\right]$}
    \put(0,73)  { 
      \begin{sideways}%
        {\small Candidates/$\left(20~\mathrm{MeV}/c^2 \right)$}
      \end{sideways}%
    }
    \put(90,105) {\small (d)  } 
    \put(90,110) {\small $\Prho^0 \to \pipi$  } 
    \put(130,110) { LHCb  } 
    \put(112,61) {\small $\mathrm{m}_{\pipi}$  }
    \put(132,61)  {\small $\left[ \mathrm{MeV}/c^2\right]$}
    \put(75,73)  { 
      \begin{sideways}%
        {\small Candidates/$\left(45~\mathrm{MeV}/c^2\right)$}
      \end{sideways}%
    }
    \put(15,45) {\small (e)  } 
    \put(15,50) {\small $\Peta^{\prime} \to \Peta\pipi$  } 
    \put(55,50) { LHCb  } 
    \put(37,1) {\small $\mathrm{m}_{\Peta\pipi}$  }
    \put(57,1)  {\small $\left[ \mathrm{MeV}/c^2\right]$}
    \put(0,13)  { 
      \begin{sideways}%
        {\small Candidates/$\left(20~\mathrm{MeV}/c^2 \right)$}
      \end{sideways}%
    }
    \put(90,45) {\small (f)  } 
    \put(90,50) {\small $\Peta \to \gamgam$  } 
    \put(130,50) { LHCb  } 
    \put(112,1) {\small $\mathrm{m}_{\gamgam}$ }
    \put(132,1)  {\small $\left[ \mathrm{MeV}/c^2\right]$}
    \put(76,13)  { 
      \begin{sideways}%
        {\small Candidates/$\left(20~\mathrm{MeV}/c^2\right)$}
      \end{sideways}%
    }
  \end{picture}
  \caption {
    Background-subtracted invariant mass distributions for 
    (a) $\gamgam$~from \mbox{$\Bs\to\jpsi\Peta(\Peta\to\gamgam)$};
    (b) $\pipipi$ from \mbox{$\Bs\to\jpsi\Peta(\Peta\to\pipipi)$};
    (c) and (d) $\pipi\g$ and $\pipi$ from \mbox{$\Bs\to\jpsi\Peta^{\prime}(\Peta^{\prime}\to\Prho^0\g,\,\Prho\to\pipi)$};
    (e) and (f) $\Peta\pipi$ and $\gamgam$ from \mbox{$\Bs\to\jpsi\Peta^{\prime}(\Peta^{\prime}\to\Peta\pipi)$}.
    The purple line is the result of the fit described in the text.
  }\label{fig:subtracted}
\end{figure}

Background-subtracted invariant mass distributions of the intermediate resonance states 
from the $\Bs\to\jpsi\mathrm{X}^0$ decays,
are shown in Fig.~\ref{fig:subtracted}. 
Clear signals are seen.
In all cases the signal yields determined from the fits are in agreement
 with the event yield in the $\Bs$ signal within one
standard deviation (Table~\ref{tab:ana_signal}).
The signal positions are consistent with the nominal
masses of the $\Peta^{(\prime)}$~mesons and the nonresonant contribution 
appears to be negligible. In each case the invariant mass resolution 
 agrees with the expectation from simulation studies.

%% file: ana3_combined.tex
\section{The $\boldsymbol{\Bd\to\jpsi\pipi}$ decay}
\label{sec:pipi}

The $\Bd\to\jpsi\Prho^0\,(\Prho^0\to\pipi)$ decay is used as a normalization
channel~\cite{Aubert:2007pipi}. Since it contains a $\jpsi$ meson and two pions 
in the final state, the systematic uncertainty is reduced
 in the ratio of the branching fractions, as the corresponding reconstruction
and particle identification uncertainties are expected to cancel.

\begin{figure}[t]
  \setlength{\unitlength}{1mm}
  \centering
  \begin{picture}(150,120)
    \put(0,0){
      \includegraphics*[width=150mm,height=120mm,%
      ]{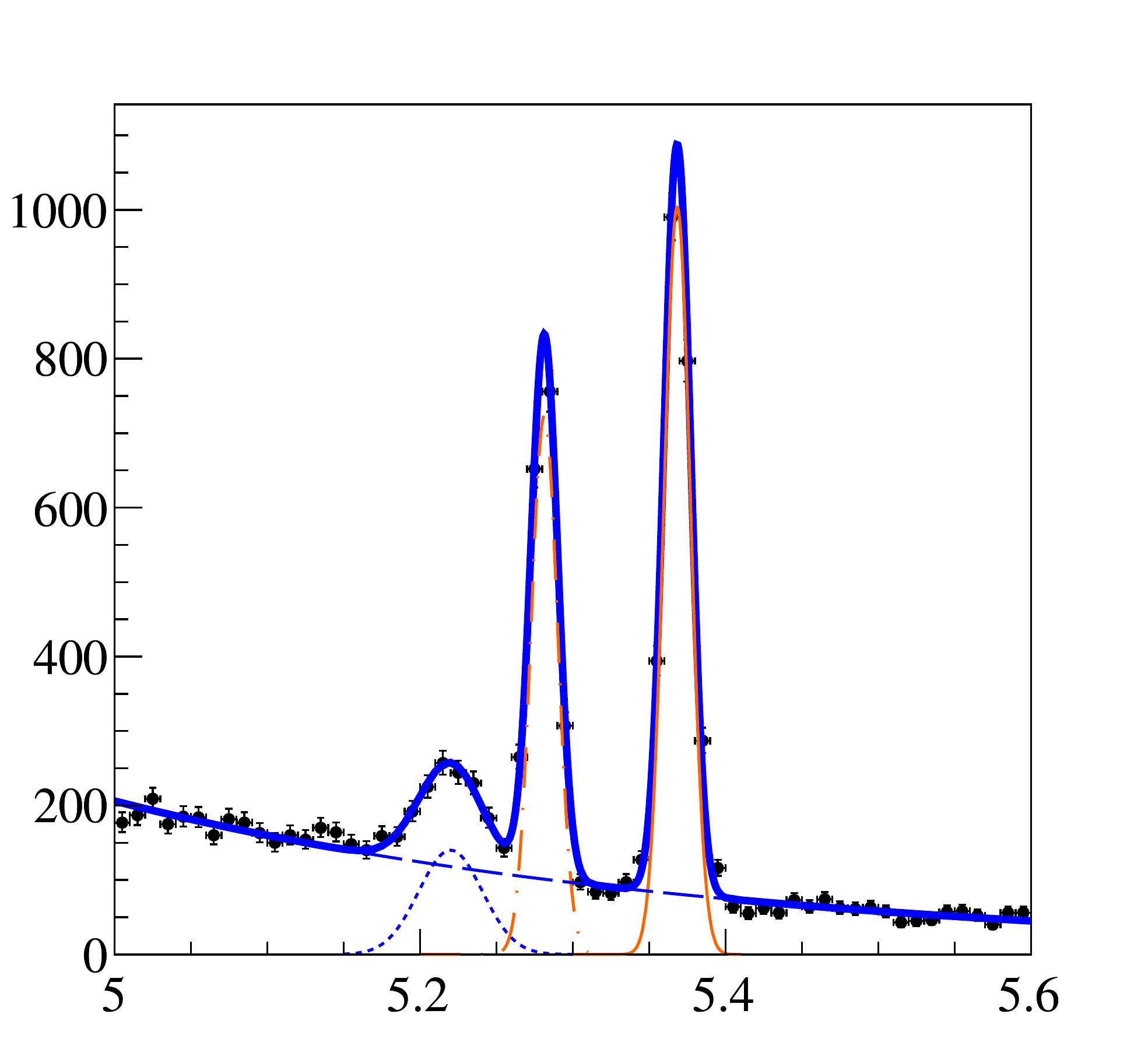}
    }
    \put(23,100) { \Large  $\mathrm{B}^0_{\mathrm{(s)}}\to\jpsi\pipi$  } 
    \put(65,-1)   { \Large $\mathrm{m}_{\jpsi\pipi}$ }
    \put(115,-1)  { \large $\left[ \mathrm{GeV}/c^2\right]$}
    \put(115,100)  { \Large{LHCb} }
    \put(-7 ,46)  { 
      \begin{sideways}%
       {\Large Candidates/$\left( 10~\mathrm{MeV}/c^2 \right)$}
      \end{sideways}%
    }
  \end{picture}
  \caption {
    Invariant mass distribution for selected $\mathrm{B}^0_{\mathrm{(s)}}\to\jpsi\pipi$~candidates.
    The black dots show the data. The dot-dashed thin orange line shows the signal $\Bd$ contribution and
    the orange solid line shows the signal $\Bs$ contribution, a reflection from misidentified
    $\Bd\to\jpsi (\mathrm{K^*}\to\mathrm{K}\Ppi)$ is shown by a blue dotted line. The blue dashed line shows the background
    contribution. The total fit function is shown as a solid blue line.
  }
  \label{fig:b_2_psi_pipi}
\end{figure}

The invariant mass spectrum for  
$\mathrm{B}^0_{\left(\mathrm{s}\right)}\to\jpsi\pipi$~candidates 
is presented in 
Fig.~\ref{fig:b_2_psi_pipi}, where
three clear signals are visible.
Two narrow signals correspond to the  
$\Bd\to\jpsi\pipi$~and 
$\Bs\to\jpsi\pipi$~decays. 
The latter decay has been studied in
detail in Refs. \cite{Aaij:2011fx, bib:bs2psibibi}.
The peak at lower mass corresponds to contamination from
 $\Bd\to\jpsi\mathrm{K}^{*0}\left(\mathrm{K}^{*0}\to\mathrm{K}^+\Ppi^-\right)$~decays
with a kaon being misreconstructed as a pion. A contribution from 
$\Bs\to\jpsi\mathrm{K}^{*0}$ decay is considered to be negligible.

The invariant mass distribution is fitted with a sum of three Gaussian functions
to describe the three signals, and an exponential function to represent the background.
The fit gives a yield of $1143\pm39$ for $\Bd\to\jpsi\pipi$.

\begin{figure}[t]
  \setlength{\unitlength}{1mm}
  \centering
  \begin{picture}(150,120)
    \put(0,0){
      \includegraphics*[width=150mm,height=120mm,%
      ]{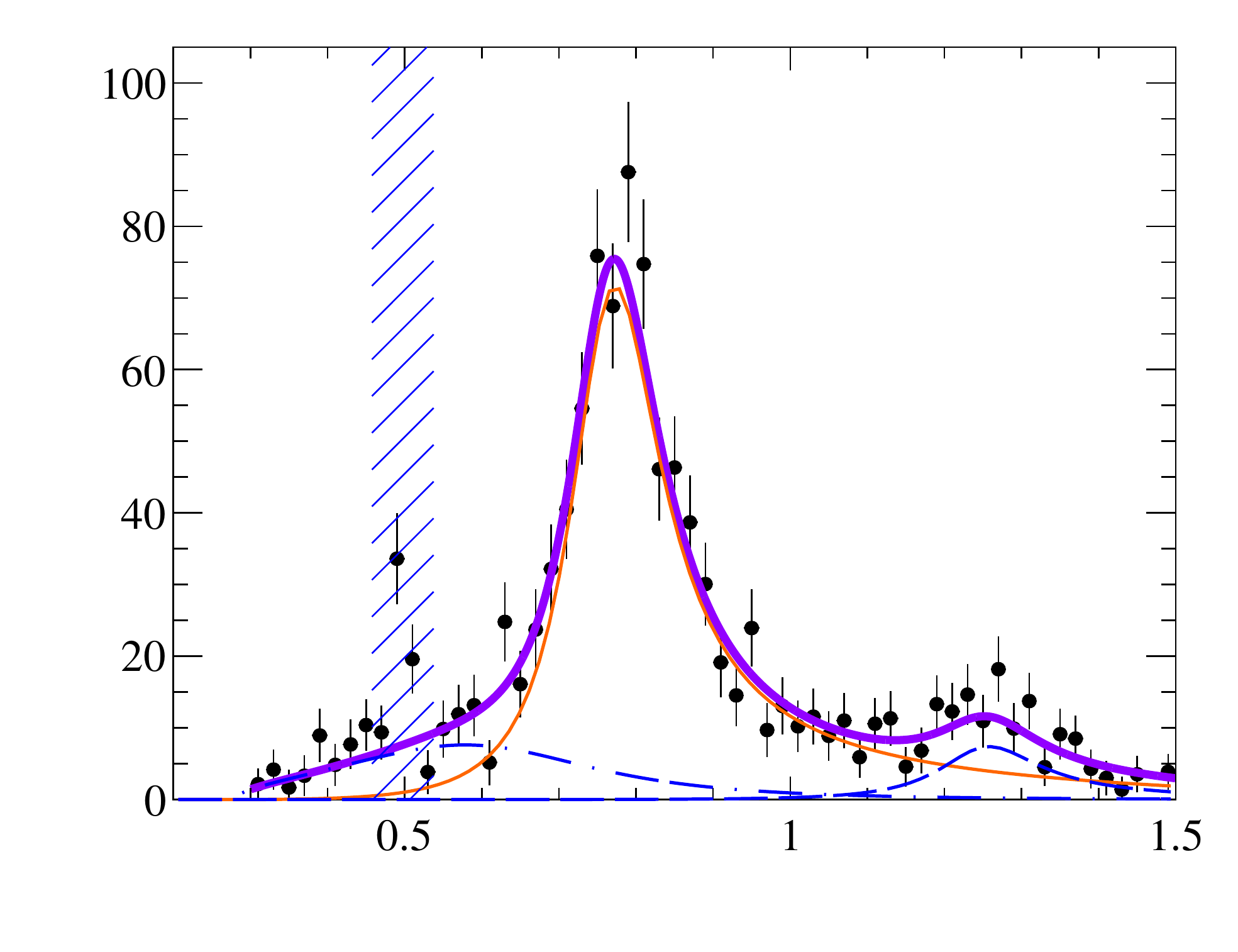}
    }
    \put(115,100)  { \Large{LHCb} }
    \put(75,6)   {\large $\mathrm{m}_{\pipi}$  }
    \put(125,6)   {\large $\left[ \mathrm{GeV}/c^2\right]$}
    \put(1 ,53)   {\Large
      \begin{sideways}%
        Candidates $\left(20~\mathrm{MeV}/c^2 \right)$
      \end{sideways}%
    }
  \end{picture}
  \caption {
    Background-subtracted \pipi~invariant mass distribution 
    from \mbox{$\Bd\to\jpsi\pipi$}~decays. The black dots show the
    data. A violet solid line denotes the total fit function,
    the solid orange line shows the $\Prho^0$ signal
    contribution and the blue dashed line shows
    the $\mathrm{f}_2(1270)$ contribution. The blue dot-dashed
    line shows the contribution from the 
    $\mathrm{f}_0(500)$. The region $\pm40~\mathrm{MeV/c}^2$ 
    around the $\mathrm{K}^0_{\mathrm{S}}$ mass is excluded
    from the fit.
  }
  \label{fig:bd_2_psi_pipi}
\end{figure}
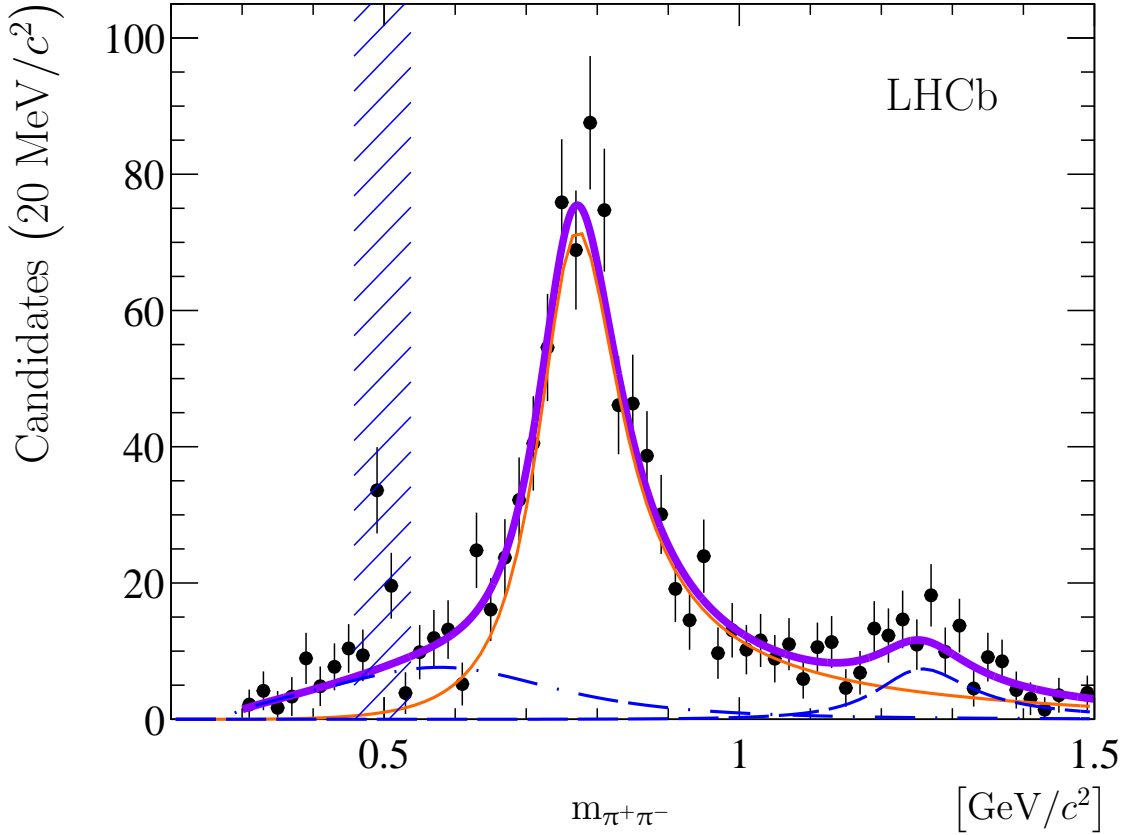

Previous studies at BaBar \cite{Aubert:2007pipi} show that 
 the $\Bd\to\jpsi\pipi$ final state has
contributions from decays of $\Prho^0$ and $\mathrm{K}^0_\mathrm{S}$
mesons, as well as a broad S-wave component. A further component
from the $\mathrm{f}_2(1270)$ resonance is also hinted at in the BaBar study. To study
the dipion mass distribution the sPlot technique is used. 
With the $\jpsi\pipi$ invariant mass as the discriminating variable, the \pipi
invariant mass spectrum from $\Bd\to\jpsi\pipi$ decays is obtained 
(see Fig.~\ref{fig:bd_2_psi_pipi}). A dominant
$\Prho^0$ signal is observed together with a narrow peak 
around 498 MeV/c$^2$ due to  $\mathrm{K}^0_{\mathrm{S}}$ decays. 
There is also a wide enhancement at a mass close to $1260~\mathrm{MeV}/c^2$. 
The position and width of this structure are consistent with the interpretation as
a contribution from the $\mathrm{f}_2(1270)$ state. 
This will be the subject of a future publication.

The distribution is fitted with the sum of several components.
A P-wave modified relativistic Breit-Wigner function \cite{Jackson:1964zd,Selleri:1962}
multiplied by a phase space factor describes the $\Prho^0$ signal.
A D-wave relativistic Breit-Wigner function is added to describe the
enhancement at $1260~\mathrm{MeV}/c^2$. The parameters (width and
mean value) of this function are fixed to the known $\mathrm{f}_2(1270)$
mass and decay width~\cite{PDG2012}. The S-wave contribution
expected from the $\mathrm{f}_0(500)$ resonance is modelled by 
a Zou-Bugg~\cite{ZouBugg:1993, Bugg:2003} function with parameters
 from Ref.~\cite{Bes:2004zb}. The $\Prho^0$ parameters (mass and width)
 are fixed at their nominal values and the region around the
 $\mathrm{K}^0_{\mathrm{S}}$ peak is excluded from
the fit. The excluded region is $\pm40~\mathrm{MeV/c}^2$
which is four times the mass resolution.
A small systematic uncertainty is induced by neglecting
the $\Prho^0 - \Pomega$ interference. The value of the
uncertainty is estimated to be 0.5\% relative to the 
$\Prho^0$ event yield.

\begin{table}[t]
  \centering
  \caption{
    Fitted yields of the $\Prho^0$~resonance, 
    the relative yields of the $\mathrm{f}_2(1270)$~and 
    $\mathrm{f}_0(500)$ components and probabilities, $\mathcal{P}$,
    of the fits to the uncorrected and efficiency-corrected
    \pipi invariant mass distributions.
  } \label{tab:bd_pipi_fit}
  \vspace*{3mm}
  \begin{small}
  \begin{tabular*}{0.9\textwidth}{@{\hspace{5mm}}l@{\extracolsep{\fill}}cc@{\hspace{5mm}}}
    & {Uncorrected fit} 
    & Efficiency-corrected fit
    \\
    \hline
    $\Prho^0$ event yield
    & $ 811    \pm 38   $\,\,\,
    & $\,\,\,\,\,\,\,(27.6  \pm 1.3 )\times10^3$
    \\
    $\mathrm{f}_0\left(500\right)$ fraction
    & $0.20 \pm 0.04$
    & $0.19 \pm 0.04$\,\,\,\,\,\,
    \\
    $\mathrm{f}_2\left(1270\right)$ fraction
    & $0.14 \pm 0.03$
    & $0.16 \pm 0.04$\,\,\,\,\,\,
    \\
    $\mathcal{P}~\left[\%\right]$
    & $40$
    & $46$\,\,\,\,\,\,
    \\
  \end{tabular*}   
  \end{small}
\end{table}

The reconstruction and selection efficiency for the
 dipion system has some dependence on the dipion invariant
 mass. A study using simulated data has shown that with the
 increase of the \pipi invariant mass in the range 300 -- 1500
 MeV/$c^2$ the efficiency decreases by approximately 16\%.
 As the $\Prho^0$ meson has a significant width, this
 dependence needs to be accounted for in the determination
 of the $\Prho^0$ signal yield. For this, the
 efficiency dependence on \pipi invariant mass extracted
 from the simulation is described with a linear
 function. Then each entry in the
 invariant mass distribution is given a weight proportional
 to the inverse value of the efficiency function and the 
efficiency-corrected invariant mass distribution is refitted with
 the same sum of functions to extract the 
efficiency-corrected event yield for $\Bd\to\jpsi\Prho^0$. 
The resulting fit parameters both for the uncorrected
and efficiency-corrected distributions are listed in Table~\ref{tab:bd_pipi_fit}.

%% file: ratios_combined.tex

\section{Measurements of ratios of branching fractions}
\label{sec:ratios}
Ratios of branching fractions are measured using the formula
\begin{equation*}
\mathcal{R}^{\mathrm{B,X}^0}_{\mathrm{B,Y}^0}\equiv
\dfrac {\BR\left(\mathrm{B}\to\jpsi \mathrm{X}^0\right)}
       {\BR\left(\mathrm{B}\to\jpsi \mathrm{Y}^0\right)} = 
\dfrac {\mathcal{Y}\left(\mathrm{B}\to\jpsi \mathrm{X}^0\right)}
       {\mathcal{Y}\left(\mathrm{B}\to\jpsi \mathrm{Y}^0\right)} 
       \times 
\dfrac {\BR_{\mathrm{Y}^0}} 
       {\BR_{\mathrm{X}^0}}
       \times 
\dfrac {\varepsilon^{\mathrm{tot}}_{\mathrm{B}\to\jpsi \mathrm{Y}^0}}
       {\varepsilon^{\mathrm{tot}}_{\mathrm{B}\to\jpsi \mathrm{X}^0}},
\end{equation*}
where $\mathcal{Y}$ are the measured event yields,
 $\varepsilon^{\mathrm{tot}}$ are the total
efficiencies, excluding the branching fractions
of light mesons and $\BR_{\mathrm{X}^0}$($\BR_{\mathrm{Y}^0}$) is the
relevant branching ratio of the light meson $\mathrm{X}^0$($\mathrm{Y}^0$) 
to the final state under consideration~\cite{PDG2012}.
In cases where decays of different types of B mesons are compared, the 
ratio of the branching fractions is multiplied by the
ratio of the corresponding b-quark hadronization fractions 
$f_\mathrm{d}/f_\mathrm{s}$~\cite{Aaij:2012hi}.

The total efficiencies consist of 
three components: the geometrical acceptance of the detector,
the reconstruction and selection efficiency and the trigger
efficiency. For the $\Bd\to\jpsi\Prho^0$ decay, the event yield 
$\mathcal{Y}$ implies the value weighted by the selection
 and reconstruction efficiency from Table~\ref{tab:bd_pipi_fit}.
Only the acceptance and trigger efficiencies are included in
$\varepsilon^{\mathrm{tot}}_{\Bd\to\jpsi\Prho^0}$.
All efficiency components have been determined using the simulation and
the values are listed in Table~\ref{tab:efficiencies}.

For channels with photons and neutral pions in the final
states, the reconstruction and selection efficiencies
are corrected for the difference in the photon
reconstruction between the data and simulation. This 
correction factor has been
determined by comparing the relative yields of the reconstructed
$\mathrm{B}^+\to\jpsi \mathrm{K}^{*+}(\mathrm{K}^{*+}\to\mathrm{K}^+\piz)$ and 
$\mathrm{B}^+\to\jpsi\mathrm{K}^+$ decays.
The results of these studies are convolved with the background
subtracted photon momentum spectra to give the correction factor
for each channel. The values of the correction factors ($\eta^{\mathrm{corr}}$)
 are also listed in Table~\ref{tab:efficiencies}.

\begin{table}[t]
  \centering
  \caption{
    Branching fractions of the intermediate resonances,
    total efficiencies (excluding the branching fractions
    of the intermediate resonances), $\varepsilon^{\mathrm{tot}}$, and 
    the photon and $\Ppi^0$ efficiency correction factors $\eta^{\mathrm{corr}}$
    for various channels. For the $\Bd\to\jpsi\Prho^0$ decay the total
    efficiency includes only the detector acceptance and trigger
    efficiencies, as the reconstruction and selection efficiency for this
    channel has been discussed in Sect.~\ref{sec:pipi}.
  } \label{tab:efficiencies}
  \vspace*{3mm}
  \begin{tabular*}{0.90\textwidth}{@{\hspace{5mm}}l@{\extracolsep{\fill}}ccc@{\hspace{5mm}}}
    Mode 
    &  $\BR~\left[ \% \right]$
    &  $\varepsilon^{\mathrm{tot}}~~\left[ \% \right]$
    &  $\eta^{\mathrm{corr}}~~\left[ \% \right]$
    \\
    \hline
    $\Bs\to\jpsi\Peta\left(\Peta\to\gamgam\right)$
    & $39.31\pm0.20$
    & $0.236 \pm 0.006 $
    & $98.0\pm7.5$ 
    \\
    $\Bs\to\jpsi\Peta\left(\Peta\to\pipipi\right)$
    & $22.74 \pm 0.28$
    & $0.059   \pm 0.002$
    & $94.1\pm7.5$
    \\
    $\Bs\to\jpsi\Peta^{\prime}\left(\Peta^{\prime}\to\Prho^0{}\g\right)$
    & $29.3 \;\; \pm 0.6\;\;$
    & $0.142   \pm 0.004$
    & $98.0\pm3.7$ 
    \\
    $ \Bs\to\jpsi\Peta^{\prime}\left(\Peta^{\prime}\to\pipi{}\Peta\right)$
    & $18.6 \;\; \pm 0.3\;\;$
    & $0.068 \pm 0.003$
    & $96.0\pm7.5$
    \\
    $\Bd\to\jpsi\Pomega\left(\Pomega\to\pipipi\right)$
    & $89.2 \;\; \pm0.7 \;\;$
    & $0.043\pm0.002$
    & $94.1\pm7.5$
    \\
    $\Bd\to\jpsi\Prho^{0}\left(\Prho^{0}\to\pipi\right)$
    & $98.90 \pm 0.16$
    & $12.6\pm0.5\;\;$
    & $\;\;$--
    \\
  \end{tabular*}   
\end{table}

\input{syst_combined}

\input{average_combined}

%% file: syst_combined.tex
\subsection{Systematic uncertainties}
\label{sec:syst}

Most systematic uncertainties cancel in the
branching fraction ratios, in particular,
 those related to the muon and
$\jpsi$ reconstruction and identification.
For the final states with photons the largest 
systematic uncertainty is related to the efficiency 
of \piz/\g~reconstruction and identification, as described above.
The uncertainties of the applied corrections 
reflect simulation statistics, and are taken
 as systematic uncertainties on the branching 
fractions ratios.
 
Another systematic uncertainty is due to
 the charged particle reconstruction efficiency which has been studied
through a comparison between data and simulation.
For the ratios where this does not cancel exactly, the corresponding
systematic uncertainty is taken to be 1.8\% per pion~\cite{bib:pi_treff}.

The systematic uncertainty related to the trigger efficiency has 
been obtained by comparison of the trigger efficiency ratios 
in data and simulation for the high yield decay mode
$\mathrm{B}^{\pm}\to\jpsi\mathrm{K}^{\pm}$ with similar kinematics 
and the same trigger requirements~\cite{Polyakov}. This uncertainty is
taken to be 1.1$\%$.

In the ratios where decays of B mesons of different types are compared ($\Bd$ or $\Bs$), 
knowledge of the hadronization fraction ratio 
$f_\mathrm{d}/f_\mathrm{s}$ is required. The measured value of this
ratio~\cite{Aaij:2012hi} has an asymmetric uncertainty of $^{+7.9}_{-7.5}\%$.

Systematic uncertainties related to the fit model
are estimated using a number of alternative models for the 
description of the invariant mass distributions.
For the \mbox{$\Bs\to\jpsi\Peta^{(\prime)}$} decays the tested alternatives 
include a fit without the \Bd component, a fit with the 
means of the Gaussians fixed to the nominal B meson masses, 
a fit with the width of the Gaussians fixed to the expected 
mass resolutions from simulation and substitution of the 
exponential background hypothesis with first- and second-order 
polynomials.
This uncertainty is calculated for the ratios of the event yields. For each 
alternative fit model the ratio of the event yields is calculated 
and the systematic uncertainty is then determined as the maximum 
deviation of this ratio from the ratio obtained with the baseline 
model.

A similar study is performed for the $\Bd\to\jpsi\Pomega$
channel. As the fit with one Gaussian function is the baseline
model in this case, here the alternative model is a fit
with two Gaussian functions (allowing a possible $\Bs$ signal).

In the $\Bd\to\jpsi\Prho^0$ case, an alternative model replaces the Zou-Bugg
$\mathrm{f}_0(500)$ term with a Breit-Wigner shape. The mass and width 
of the broad $\mathrm{f}_0(500)$ state are not well known. The mass measured by various
experiments varies in a range between 400 and \mbox{1200 $\mathrm{MeV}/c^2$} and 
the measured width ranges between 600 and 1000 $\mathrm{MeV}/c^2$~\cite{PDG2012}. 
Therefore, the $\mathrm{f}_0(500)$
parameters are varied in this range and the $\Prho^0$ yield
is determined. Again, the maximum deviation from the baseline 
model is treated as the systematic uncertainty of the fit.

The uncertainties related to the knowledge of the branching fractions 
of~$\Peta$, $\Peta^{\prime}$, $\piz$~and $\Pomega$~decays are 
taken from Ref.~\cite{PDG2012}. Other
systematic uncertainties, such as those related to the selection criteria 
are negligible.
The systematic uncertainties are summarized in Tables~\ref{tab:syst1}
and~\ref{tab:syst2}. The total systematic 
uncertainties are estimated using a simulation technique (see Sect.~\ref{seq:average}).

\begin{table}[t]
  \centering
  \caption{
    Relative systematic uncertainties for 
    ratios of the branching fractions ($\mathcal{R}$) for the
    $\Bs\to\jpsi\Peta^{(\prime)}$ channels $\left[\%\right]$.
  } \label{tab:syst1}
  \vspace*{3mm}
  \begin{tabular*}{1.00\textwidth}{@{\hspace{0mm}}l@{\extracolsep{\fill}}cccccc@{\hspace{0mm}}}
    Parameter
    & $\mathcal{R}^{\Peta\to\gamgam}_{\Peta\to\pipipi}$ 
    & $\mathcal{R}^{\Peta^{\prime}\to\Peta\pipi}_{\Peta\to\gamgam}$ 
    & $\mathcal{R}^{\Peta^{\prime}\to\Peta\pipi}_{\Peta\to\pipipi}$
    & $\mathcal{R}^{\Peta^{\prime}\to\Prho^0\gamma}_{\Peta\to\gamgam}$ 
    & $\mathcal{R}^{\Peta^{\prime}\to\Prho^0\gamma}_{\Peta\to\pipipi}$
    & $\mathcal{R}^{\Peta^{\prime}\to\Prho^0\gamma}_{\Peta^{\prime}\to\Peta\pipi}$
    \\
    \hline 
    $\Peta_{\mathrm{corr}}$
    & --
    & --
    & --
    & $3.8$
    & $3.9$
    & $3.9$
    \\
    $\Ppi^{\pm}$~reco
    & $2\times1.8$ 
    & $2\times1.8$
    & --  
    & $2\times1.8$
    & --  
    & --  
    \\
    Trigger
    & 1.1
    & 1.1
    & 1.1 
    & 1.1
    & 1.1
    & 1.1
    \\
    Fit function
    & $^{+3.7}_{-0.0}$ \:\: 
    & $^{+9.9}_{-0.0}$ \:\:
    & $^{+1.3}_{-5.6}$ \:
    & $^{+3.4}_{-0.0}$ \:\:
    & $<0.1$ \;\;
    & $^{+0.0}_{-2.8}$ \;\;
    \\
    $\BR\left( \Peta, \Peta^{\prime},\Pomega\right)$
    & 1.3
    & 1.7
    & 2.0
    & 2.1
    & 1.8
    & 2.6
    \\
  \end{tabular*}   
\end{table}

\begin{table}[t]
  \centering
  \caption{
    Systematic uncertainties for 
    ratios of the branching fractions ($\mathcal{R}$)
    relative to $\Bd\to\jpsi\Prho^0$ $\left[\%\right]$.
  } \label{tab:syst2}
  \vspace*{3mm}
  \begin{tabular*}{1.\textwidth}{@{\hspace{0mm}}l@{\extracolsep{\fill}}ccccc@{\hspace{0mm}}}
    Parameter 
    & $\mathcal{R}^{\Bs,\Peta\to\gamgam}_{\Bd,\Prho^0\to\pipi}$ 
    & $\mathcal{R}^{\Bs,\Peta\to\pipipi}_{\Bd,\Prho^0\to\pipi}$ 
    & $\mathcal{R}^{\Bs,\Peta^{\prime}\to\Prho^0\gamma}_{\Bd,\Prho^0\to\pipi}$ 
    & $\mathcal{R}^{\Bs,\Peta^{\prime}\to\Peta\pipi}_{\Bd,\Prho^0\to\pipi}$
    & $\mathcal{R}^{\Bd,\Pomega\to\pipipi}_{\Bd,\Prho^0\to\pipi}$
    \\
    \hline 
    $\Peta_{\mathrm{corr}}$
    & $7.6$
    & $8.0$
    & $3.8$
    & $7.8$
    & $8.0$
    \\
    $\Ppi^{\pm}$~reco
    & $2\times1.8$ 
    & --
    & --
    & --  
    & --  
    \\
    Trigger
    & 1.1
    & 1.1
    & 1.1 
    & 1.1
    & 1.1
    \\
    Fit function
    & $^{+5.1}_{-3.7}$ \:
    & $^{+5.0}_{-4.3}$ \:
    & $^{+5.0}_{-5.7}$ \:
    & $^{+5.0}_{-8.7}$ \:
    & $^{+6.4}_{-8.8}$ \:
    \\
    $\BR\left( \Peta, \Peta^{\prime},\Pomega\right)$
    & 0.5
    & 1.2
    & 2.1
    & 1.6
    & 0.8
    \\
  \end{tabular*}   
\end{table}

%% file: average_combined.tex
\subsection{Results}
\label{seq:average}

The final ratios
$\mathcal{R}^{\Bs,\Peta^{\prime}}_{\Bs,\Peta}$,
$\mathcal{R}^{\Bs,\Peta^{\left(\prime\right)}}_{\Bd,\Prho^0}$ and 
$\mathcal{R}^{\Bd,\Pomega}_{\Bd,\Prho^0}$ are determined 
using a procedure that combines
$\chi^2$-minimization with constraints and 
simplified simulation. 
First, the $\chi^2$ is minimized
\begin{equation}
\chi^2 = \sum_i \chi^2_i, \nonumber
\label{eq:chi2}
\end{equation}
where the sum is performed over the six measured event yields for the six 
different modes:
$\Bs\to\jpsi\Peta\left( \Peta\to\gamgam\right)$,
$\Bs\to\jpsi\Peta\left( \Peta\to\pipi{}\pi^0\right)$,
$\Bs\to\jpsi\Peta^{\prime}\left( \Peta^{\prime}\to\Prho^0{}\g\right)$,
\mbox{$\Bs\to\jpsi\Peta^{\prime}\left( \Peta^{\prime}\to\Peta{}\pipi\right)$},
$\Bd\to\jpsi\Pomega$ and 
$\Bd\to\jpsi\Prho^0$, and 
$\chi^2_i = \frac{ \left( x - \mathcal{Y}_i \right)^2}{ \sigma^2_{\mathcal{Y}_i}}$.
In this procedure the following constraints are imposed
\begin{subequations}
\begin{eqnarray}
\dfrac{\mathcal{Y}_{\Bs\to\jpsi\Peta\left( \Peta\to\gamgam\right)}} 
      {\varepsilon_{\Bs\to\jpsi\Peta\left( \Peta\to\gamgam\right)}\times\BR\left( \Peta\to\gamgam\right)}  & = & 
\dfrac{\mathcal{Y}_{\Bs\to\jpsi\Peta\left( \Peta\to\pipi{}\Ppi^0\right)}} 
      {\varepsilon_{\Bs\to\jpsi\Peta\left( \Peta\to\pipi{}\Ppi^0\right)}\times\BR\left( \Peta\to\pipi{}\Ppi^0\right)}, \nonumber \\ 
\dfrac{\mathcal{Y}_{\Bs\to\jpsi\Peta^{\prime}\left( \Peta^{\prime}\to\Prho^0{}\g\right)}} 
      {\varepsilon_{\Bs\to\jpsi\Peta^{\prime}\left( \Peta^{\prime}\to\Prho^0{}\g\right)}\times\BR\left( \Peta^{\prime}\to\Prho^0{}\g\right)}  & = & 
\dfrac{\mathcal{Y}_{\Bs\to\jpsi\Peta^{\prime}\left( \Peta^{\prime}\to\Peta{}\pipi\right)}} 
      {\varepsilon_{\Bs\to\jpsi\Peta^{\prime}\left( \Peta^{\prime}\to\Peta{}\pipi\right)}\times\BR\left( \Peta^{\prime}\to\Peta{}\pipi\right)}.\nonumber  
\end{eqnarray}
\label{eq:lambda}
\end{subequations}
The ratios 
$\mathcal{R}^{\Bs,\Peta^{\prime}}_{\Bs,\Peta}$,
$\mathcal{R}^{\Bs,\Peta^{\left(\prime\right)}}_{\Bd,\Prho^0}$ and 
$\mathcal{R}^{\Bd,\Pomega}_{\Bd,\Prho^0}$ are
determined using the event yields obtained from 
the minimization procedure. 
For this determination the efficiencies 
$\varepsilon_i$ have been varied using a
simplified simulation taking into account correlations 
between the various components where appropriate.
As both the $\chi^2$ and 
the ratios $\mathcal{R}$ depend only on 
the ratios of efficiencies, systematic uncertainties
are minimized.
The remaining systematic 
uncertainties have been taken into account
as uncertainties in the efficiency ratios.
In total, $10^6$ simulated experiments with 
 different settings of $\varepsilon_i$ have been performed.
The symmetric 68\% intervals have been assigned
as the systematic uncertainty.

The obtained ratios $\mathcal{R}$ are
\begin{subequations}
\begin{eqnarray*}
\mathcal{R}^{\Bs,\Peta^{\prime}}_{\Bs,\Peta}        & = & 0.90 \pm 0.09\,^{+0.06}_{-0.02}, \\ 
\mathcal{R}^{\Bs,\Peta}_{\Bd,\Prho^0}          & = & 3.75  \pm 0.31\,^{+0.30}_{-0.40} \times\left(\frac{f_\mathrm{d}}{f_\mathrm{s}}\right), \\ 
\mathcal{R}^{\Bs,\Peta^{\prime}}_{\Bd,\Prho^0}      & = & 3.38 \pm  0.30\,^{+0.14}_{-0.36} \times\left(\frac{f_\mathrm{d}}{f_\mathrm{s}}\right), \\ 
\mathcal{R}^{\Bd,\Pomega}_{\Bd,\Prho^0}      & = & 0.89\pm0.19\,^{+0.07}_{-0.13}, 
\end{eqnarray*}
\end{subequations}
where the first uncertainty is statistical and the second is systematic.

%% file: summary_combined.tex
\section{Summary}
\label{sec:summ}
With 1.0 fb$^{-1}$ of data, collected in 2011 with the LHCb detector,
 the first evidence for the $\Bd\to\jpsi\Pomega$
decay has been found, and its branching fraction,
normalized to that of 
the $\Bd\to\jpsi\Prho^0$ decay, is measured to be

$$\dfrac{\BR(\Bd\to\jpsi\Pomega)}{\BR(\Bd\to\jpsi\Prho^0)} = 
0.89\pm0.19\,(\mathrm{stat})\,^{+0.07}_{-0.13}\,(\mathrm{syst}).$$

\noindent Multiplying by the known value of 
$\BR(\Bd\to\jpsi\Prho^0) = (2.7\pm0.4)\times10^{-5}$~\cite{Aubert:2007pipi}, the absolute value of the 
branching fraction is
$$
\BR(\Bd\to\jpsi\Pomega) = (2.41\pm0.52\,(\mathrm{stat})\,^{+0.19}_{-0.35}\,(\mathrm{syst})\pm0.36\,(\BR_{\Bd\to\jpsi\Prho^0}))\times10^{-5}.
$$

Using the same dataset, the ratio of
the branching fractions of $\Bs\to\jpsi\Peta$ and
$\Bs\to\jpsi\Peta^{\prime}$ decays
has been measured. As each of the decays has been
reconstructed in two final states, the resulting ratio has been 
calculated through an averaging procedure to be
\begin{equation*}
\mathcal{R}^{\Bs,\Peta^{\prime}}_{\Bs,\Peta} = 
\dfrac{\BR(\Bs\to\jpsi\Peta^{\prime})}{\BR(\Bs\to\jpsi\Peta)} = 
0.90\pm0.09\,(\mathrm{stat})\,^{+0.06}_{-0.02}\,(\mathrm{syst}).
\end{equation*}
This result is consistent with the previous
 Belle measurement of
$\mathcal{R}^{\Bs,\Peta^{\prime}}_{\Bs,\Peta}~=~0.73~\pm~0.14$~\cite{Adachi:2009usa},
but is more precise. Assuming that the contribution
from the purely gluonic component is negligible, this ratio
corresponds to a value of the $\Peta - \Peta^{\prime}$ mixing phase of 
$\Pphi_\mathrm{P} = \left(45.5\,^{+1.8}_{-1.5}\right)^\circ$.
The branching fractions of the $\Bs\to\jpsi\Peta$
 and $\Bs\to\jpsi\Peta^{\prime}$ decays have been
determined by normalization to the 
$\Bd\to\jpsi\Prho^0$ decay branching
fraction, and 
using the known value of $f_\mathrm{s}/f_\mathrm{d} = 0.267\,^{+0.021}_{-0.020}$~
\cite{Aaij:2012hi} their ratios are 
\begin{eqnarray*}
\frac{\BR(\Bs\to\jpsi\Peta)}{\BR(\Bd\to\jpsi\Prho^0)} &=& \:14.0 
\pm1.2\,(\mathrm{stat})\,^{+1.1}_{-1.5}\,(\mathrm{syst})\,^{+1.1}_{-1.0}\left(\frac{f_\mathrm{d}}{f_\mathrm{s}}\right),\\
\frac{\BR(\Bs\to\jpsi\Peta^{\prime})}{\BR(\Bd\to\jpsi\Prho^0)} &=& 
\:12.7\pm1.1\,(\mathrm{stat})\,^{+0.5}_{-1.3}\,(\mathrm{syst})\,^{+1.0}_{-0.9}\left(\frac{f_\mathrm{d}}{f_\mathrm{s}}\right).
\end{eqnarray*}
When multiplying by the known value of $\BR(\Bd\to\jpsi\Prho^0)$,
the branching fractions are measured as
\begin{eqnarray*}
\BR(\Bs\to\jpsi\Peta) & = & \left(3.79\pm0.31\,(\mathrm{stat})\,^{+0.20}_{-0.41}\,(\mathrm{syst})\,^{+0.29}_{-0.27}\left(\tfrac{f_\mathrm{d}}{f_\mathrm{s}}\right)\pm0.56\,(\BR_{\Bd\to\jpsi\Prho^0})\right)\times10^{-4},\\ 
\BR(\Bs\to\jpsi\Peta^{\prime}) & = & \left(3.42\pm0.30\,(\mathrm{stat})\,^{+0.14}_{-0.35}\,(\mathrm{syst})\,^{+0.26}_{-0.25}\left(\tfrac{f_\mathrm{d}}{f_\mathrm{s}}\right)\pm0.51\,(\BR_{\Bd\to\jpsi\Prho^0})\right)\times10^{-4}.
\end{eqnarray*}

The branching fractions measured here correspond to the time integrated quantities, while theory predictions usually refer to the 
branching fractions at $t = 0$. Special care needs to be taken when the $\Bs$ and $\Bd$ decays are compared at the amplitude level,
corresponding to the branching ratio at $t = 0$~\cite{DeBruyn:2012}. Since the $\jpsi \Peta^{(\prime)}$ final states are \CP-eigenstates, the size of 
this effect can be as large as 10\%, and can be corrected for using input from theory or determined from effective lifetime 
measurements~\cite{DeBruyn:2012}. With a larger dataset such measurements, as well as studies of $\Peta-\Peta^{\prime}$ mixing and measurements of \CP asymmetries 
in the $\Bs \to \jpsi\Peta^{(\prime)}$ modes will be possible.

%% file: acknowledgements.tex
\section*{Acknowledgements}

\noindent We would like to thank A.K. Likhoded for many fruitful discussions.
We express our gratitude to our colleagues in the CERN accelerator
departments for the excellent performance of the LHC. We thank the
technical and administrative staff at CERN and at the LHCb institutes,
and acknowledge support from the National Agencies: CAPES, CNPq,
FAPERJ and FINEP (Brazil); CERN; NSFC (China); CNRS/IN2P3 (France);
BMBF, DFG, HGF and MPG (Germany); SFI (Ireland); INFN (Italy); FOM and
NWO (The Netherlands); SCSR (Poland); ANCS (Romania); MinES of Russia and
Rosatom (Russia); MICINN, XuntaGal and GENCAT (Spain); SNSF and SER
(Switzerland); NAS Ukraine (Ukraine); STFC (United Kingdom); NSF
(USA). We also acknowledge the support received from the ERC under FP7
and the Region Auvergne.